\documentclass[aps,pra,final,10pt,twocolumn,superscriptaddress]{revtex4-1}
\usepackage{amsmath,amssymb,bm}
\usepackage{graphicx,color}
\usepackage[squaren]{SIunits}
\usepackage[english]{babel}
\usepackage{amstext}
\usepackage{amsthm}
\usepackage{latexsym}
\usepackage{array}
\usepackage{color}
\usepackage{float}
\usepackage{microtype}
 
\usepackage{multirow}
\usepackage{dcolumn}
\usepackage{soul}
\usepackage{lettrine}
\usepackage{type1cm} 
\usepackage{natbib}
\usepackage{lineno}

\definecolor{url}{RGB}{0,20,160}
\usepackage[colorlinks=true,linkcolor=blue,citecolor=blue,urlcolor=url]{hyperref}
\usepackage[usenames,dvipsnames,svgnaes,table]{xcolor}
\makeatletter
\def\frutiger{cmss10 }
\def\frutigerbold{cmssbx10 }
=\frutigerbold at 14pt
=\frutiger at 8pt
=\frutigerbold at 12pt
=\frutigerbold at10pt
=\frutigerbold at 8pt
=\frutiger at 8pt
=\frutigerbold at 31pt
\def\@caption@tabnum@sep{\figtextfont{{ }{\bf\textbar}{ }}}%
\def\fnum@table{{\bf\tablename~\thetable}}

\def\@caption@fignum@sep{\figtextfont{{ }{\bf\textbar}{ }}}%
\def\fnum@figure{{\bf\figurename~\thefigure}}
\renewenvironment{figure}{\@float{figure}\def\textbf##1{{\fignumfont ##1}}\def\bf{\fignumfont}}{\end@float}
\def\@startsection#1#2#3#4#5#6{%
	\if@noskipsec\leavevmode\fi
	\par\@tempskipa #4\relax
	\@afterindenttrue
	\ifdim\@tempskipa <\z@
	\@tempskipa -\@tempskipa \@afterindentfalse
	\fi\if@nobreak\everypar{}%
	\else\addpenalty\@secpenalty\addvspace\@tempskipa\fi
	\@ifstar{\@ssect{#3}{#4}{#5}{#6}}{\@dblarg{\@sect{#1}{#2}{#3}{#4}{#5}{#6}}}}
\def\@sect#1#2#3#4#5#6[#7]#8{%
	\ifnum #2>0
	\let\@svsec\@empty
	\else\refstepcounter{#1}\protected@edef\@svsec{\@seccntformat{#1}\relax}\fi
	\@tempskipa #5\relax
	\ifdim\@tempskipa>\z@
	\begingroup#6{\@hangfrom{\hskip #3\relax\@svsec}%
		\interlinepenalty \@M #8\@@par}\endgroup
	\csname #1mark\endcsname{#7}%
	\addcontentsline{toc}{#1}{%
		\ifnum #2>\c@secnumdepth\else
		\protect\numberline{\csname the#1\endcsname}\fi #7}%
	\else\def\@svsechd{#6{\hskip #3\relax
			\@svsec #8\ifnum#2=2.\fi}%
		\csname #1mark\endcsname{#7}%
		\addcontentsline{toc}{#1}{%
			\ifnum #2>\c@secnumdepth \else
			\protect\numberline{\csname the#1\endcsname}\fi #7}}%
	\fi\@xsect{#5}}

\renewcommand\section{\@startsection {section}{1}{\z@}%
	{-10pt \@plus -1ex \@minus -.2ex}{.5ex }{\normalfont\Large\bfseries\sectionfont}}
\renewcommand\subsection{\@startsection{subsection}{2}{\z@}%
	{10pt\@plus 1ex \@minus .2ex}{-0.5ex \@plus .2ex}{\normalfont\large\bfseries\subsectionfont}}
\def\frontmatter@title@format{\titlefont\centering}%
\def\frontmatter@title@below{\addvspace{-5pt}}%

\renewcommand\NAT@biblabelnum[1]{#1.}
\renewcommand\NAT@citesuper[3]{\ifNAT@swa
\unskip\hspace{1\p@}\textsuperscript{(#1)}%
   \if\relax#3\relax\else\ (#3)\fi\else (#1)\fi\endgroup}

\newcommand*\bib@heading{%
	\section{\refname}
	\fontsize{8}{10}\selectfont
}
\newcommand*\@openbib@code{%
	\advance\leftmargin\bibindent
	\itemindent -\bibindent
	\listparindent \itemindent
	\parsep \z@
}%
\newdimen\bibindent
\usepackage[usenames,dvipsnames,svgnaes,table]{xcolor}
\definecolor{col1}{rgb}{0.0, 0.30, 1.0}
\definecolor{col2}{rgb}{0.9, 0.0, 0.30}

\usepackage[draft]{changes} 
\usepackage[normalem]{ulem}
\definechangesauthor[name={yi}, color=red]{Yi}

\makeatletter

\newcommand{\Rmnum}[1]{\expandafter\@slowromancap\romannumeral #1@}
\newcommand{\revise}[1]{{\textcolor{red}}}

\begin{document}
\title{Exploring low lattice thermal conductivity materials using chemical bonding principles}

\author{Jiangang He}
\email{jiangang2020@gmail.com}
\affiliation{Department of Materials Science and Engineering, Northwestern University, Evanston, IL 60208, USA}

\author{Yi Xia}
\altaffiliation{These authors contributed equally: Jiangang He, Yi Xia}
\affiliation{Department of Materials Science and Engineering, Northwestern University, Evanston, IL 60208, USA}

\author{Wenwen Lin}
\affiliation{Materials Science Division, Argonne National Laboratory, Argonne, IL 60439, USA}

\author{Koushik Pal}
\affiliation{Department of Materials Science and Engineering, Northwestern University, Evanston, IL 60208, USA}

\author{Yizhou Zhu}
\affiliation{Department of Materials Science and Engineering, Northwestern University, Evanston, IL 60208, USA}

\author{Mercouri G. Kanatzidis}
\affiliation{Materials Science Division, Argonne National Laboratory, Argonne, IL 60439, USA}
\affiliation{Department of Chemistry, Northwestern University, Evanston, IL 60208, USA}

\author{Chris  Wolverton}
\email{c-wolverton@northwestern.edu}
\affiliation{Department of Materials Science and Engineering, Northwestern University, Evanston, IL 60208, USA}

\date{\today}

\begin{abstract}
\noindent
Semiconductors with very low lattice thermal conductivities are highly desired for applications relevant to thermal energy conversion and management, such as thermoelectrics and thermal barrier coatings. Although the crystal structure and chemical bonding are known to play vital roles in shaping heat transfer behavior, material design approaches of lowering lattice thermal conductivity using chemical bonding principles are uncommon. In this work, we present an effective strategy of weakening interatomic interactions and therefore suppressing lattice thermal conductivity based on chemical bonding principles and develop a high-efficiency approach of discovering low $\kappa_{\rm L}$ materials by screening the local coordination environments of crystalline compounds. The followed first-principles calculations uncover 30 hitherto unexplored compounds with (ultra)low lattice thermal conductivities from thirteen prototype crystal structures contained in the inorganic crystal structure database. Furthermore, we demonstrate an approach of rationally designing high-performance thermoelectrics by additionally incorporating cations with stereochemically active lone-pair electrons. Our results not only provide fundamental insights into the physical origin of the low lattice thermal conductivity in a large family of copper-based compounds but also offer an efficient approach to discovery and design materials with targeted thermal transport properties.
\end{abstract}

\maketitle

Crystalline solids with very low lattice thermal conductivities ($\kappa_{\rm L}$) are both fundamentally interesting and technologically important in generating, converting, and managing thermal energy\cite{10005741159,darolia2013thermal,wehmeyer2017thermal}. In the simple kinetic theory\cite{tritt2005thermal}, $\kappa_{\rm L} = \frac{1}{3}Cv_g^2\tau$, where $C$, $v_g$, and $\tau$ are heat capacity, phonon group velocity ($v_g$), and phonon relaxation time ($\tau$), respectively. Therefore, materials with low $v_g$, short $\tau$, or small $C$ usually have low $\kappa_{\rm L}$. Decades studies on thermoelectric materials have explored many strategies of designing and discovering new materials with low $\kappa_{\rm L}$, as well as minimizing $\kappa_{\rm L}$ of the well-studied materials\cite{toberer2011phonon,mao2018advances,chen2018manipulation}. The widely-used approaches that can effectively shorten $\tau$ are based on introducing defects\cite{mao2018advances}, nano-structure precipitates\cite{hsu2004cubic,biswas2012high,poudel2008high,pei2011high}, lone pair electrons ions\cite{PhysRevLett.107.235901,nielsen2013lone}, rattling phonon modes\cite{PhysRevLett.114.095501,tritt2001recent,nolas2001semiconductor,lin2016concerted,he2016ultralow}, and ferroelectric instability-induced phonon softening\cite{delaire2011giant,sarkar2020ferroelectric}.
Since $v_g$ is proportional to $\sqrt{k/M}$, where $k$ and $M$ are respectively the bond stiffness (or bond strength) and the atomic mass\cite{tritt2005thermal}, low $v_g$ are also expected in materials with weak chemical bonds and large atomic mass. Taking advantage of the large atomic mass is straightforward and has been utilized to screen new thermoelectric materials with low $\kappa_{\rm L}$\cite{wang2020cu3erte3}. However, bond strength is more complicated and correlated to the electronegativities of constituting atoms and local coordination of a crystal structure. Generally, the bond in the environment with a higher coordination number (CN) is weaker than that in a lower CN one due to the longer bond length between cation and anion in the high CN case, reflecting Pauling's second rule\cite{pauling1929principles}. For example, the bond length in rock-salt structure (octahedral coordination, CN=6) is usually longer than that in the zinc blende structure (tetrahedral coordination, CN=4). As a consequent, the rock-salt compounds (e.g., NaCl: 7.1 Wm$^{-1}$K$^{-1}$, RbCl: 2.8 Wm$^{-1}$K$^{-1}$ at 300 K \cite{shinde2006high}) usually have much lower $\kappa_{\rm L}$ than the zinc blende ones (e.g., GaAs: 45 Wm$^{-1}$K$^{-1}$, ZnSe: 19 Wm$^{-1}$K$^{-1}$ at 300 K)\cite{spitzer1970lattice,zeier2016thinking,morelli2006high}. On the other hand, since the wurtzite-type structure has the same local coordination with the zinc blende one (the difference is the stacking ordering), it has very similar $\kappa_{\rm L}$ as zinc blende\cite{PhysRevB.91.094306,morelli2006high}.

Surprisingly, a few copper-based compounds with the zinc blende structure and small average atomic masses ($\overline{M}$ = 49 $\sim$ 95) show unexpected low $\kappa_{\rm L}$. For instance, the $\kappa_{\rm L}$ of CuCl, CuBr, and CuI are 0.84, 1.25, 1.68 Wm$^{-1}$K$^{-1}$ at 300 K, respectively\cite{perry2011handbook,PhysRevB.26.1873}, which are even lower than the heavier ($\overline{M}$ = 167) and strongly anharmonic compound PbTe\cite{morelli2008intrinsically,delaire2011giant}. 
Compared to the conventional zinc blende compounds, these compounds display unique lattice dynamics properties, featuring low-frequency dispersions and soft optical modes, which have been attributed to the large ionic radius mismatch between Cu$^{+}$ and Cl$^{-}$ ions\cite{PhysRevB.96.100301}. Meanwhile, experiments have shown a large thermal displacement parameter of Cu in CuCl\cite{hull1996superionic}, indicating a weak Cu-Cl bond and corroborating the existence of low-frequency phonons. Another set of interesting compounds include $\beta$-BaCu$_2$S$_2$, which crystallizes in the ThCr$_2$Si$_2$-type structure and possesses an ultralow $\kappa_{\rm L}$ of 0.68 Wm$^{-1}$K$^{-1}$ at 300 K\cite{kurosaki2005thermoelectric} and BiCuSeO, which adopts ZrCuSiAs-type structure and has a $\kappa_{\rm L}$ of 1.0 Wm$^{-1}$K$^{-1}$ at 300 K\cite{zhao2010bi}. The common feature of these two structures is the presence of edge-sharing Cu$X_4$ ($X$=S and Se) tetrahedron. It is interesting to note that all the listed compounds contain Cu, which, akin to Ag, generally forms superionic conductors at high temperatures due to the weak chemical bond between Cu$^{+}$/Ag$^{+}$ and anions\cite{hull1996superionic,miller2013mechanism}. Considering the facts that (i) Cu is an abundant earth element and the synthesis of Cu-based compounds is relatively simple, and (ii) Cu-based thermoelectric materials have a long research history back to 1827, and many of them have good thermoelectric performance\cite{Becauerel1827,qiu2016cu,zhao2010bi,zhao2014bicuseo,liu2012copper,shi2009thermoelectric,nunna2017ultrahigh,qiu2018suppression,wei2019copper}, it is intriguing to ask what makes the thermal transport behaviors of these compounds so anomalous?

These observations motivate us to investigate the origin of low $\kappa_{\rm L}$ and its relation to the distinctive chemical bonds in Cu/Ag-based compounds and to ultimately search for new materials with very low $\kappa_{\rm L}$ by exploiting the underlying mechanisms. We find that the intrinsically low $\kappa_{\rm L}$ of these compounds is mainly attributable to the weak bonds between Cu/Ag and anions, originating from the antibonding interactions between Cu/Ag-$d$ orbitals and anion-$p$ orbitals through an unusual $p$-$d$ coupling. Moreover, we demonstrate that the antibonding interaction can be further weakened by introducing edge-sharing and face-sharing polyhedra centered at Cu$^{+}$/Ag$^{+}$, in the spirit of Pauling's third rule\cite{pauling1929principles}. Based on these new findings, we then perform a comprehensive screening for compounds exhibiting these local coordination environments within the inorganic crystal structure database (ICSD). We discovered 30 hitherto unexplored compounds that potentially exhibit very low $\kappa_{\rm L}$, as evidenced by our first-principles anharmonic lattice dynamics simulations. To make these compounds suitable for thermoelectric applications, we further propose a general strategy of improving electrical transport properties by incorporating cations with stereochemically active lone-pair electrons. The materials design philosophy based on the crystal and structural chemistry unveiled by this study is universal, and hence should aid the rational design of thermal management materials and thermoelectric materials.

\begin{figure}
	\centering
	\includegraphics[width=1\linewidth]{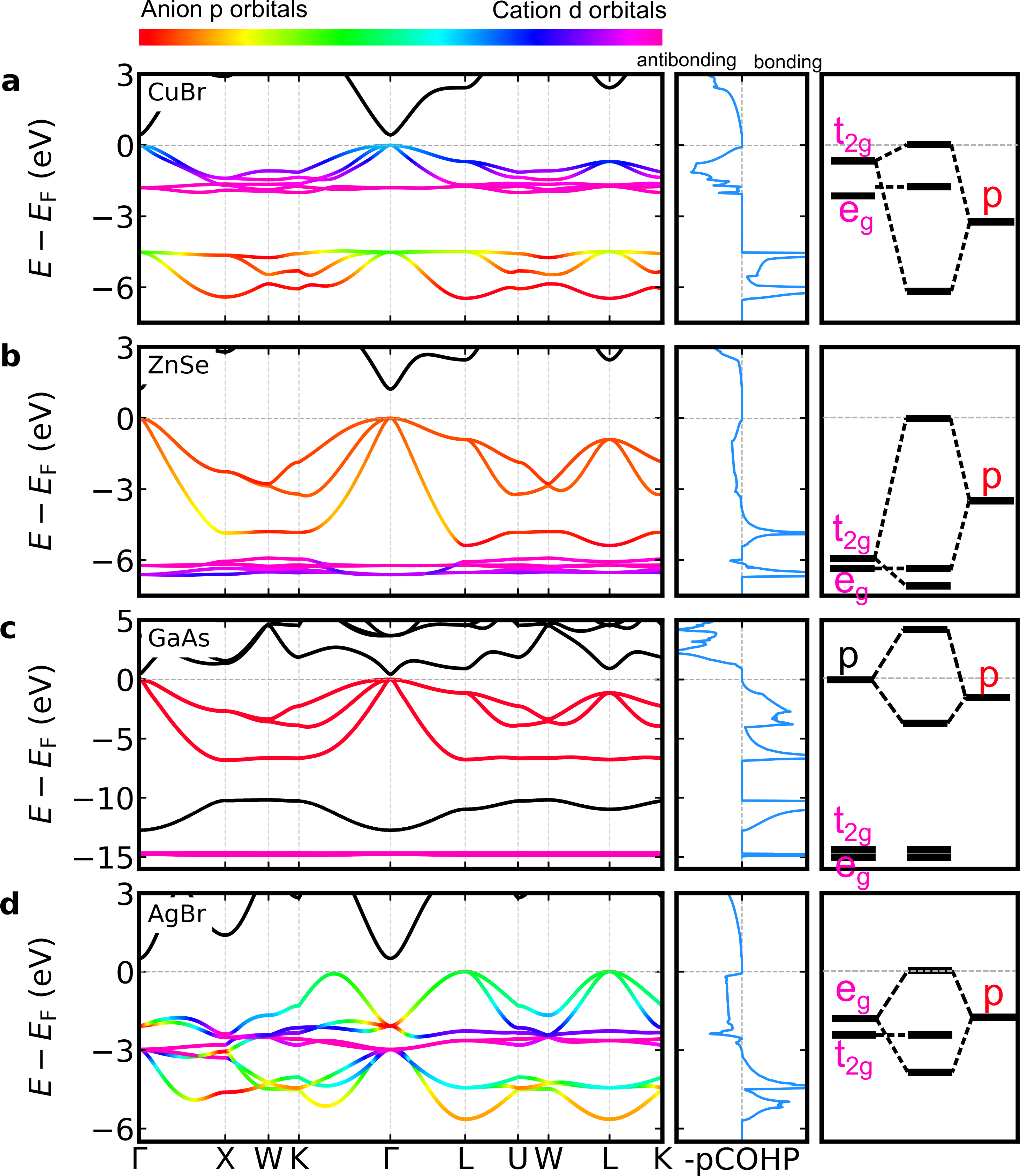}
	\caption{Orbital-projected band structures, -pCOHP, and molecular orbital diagrams of binary compounds. \textbf{a} zinc blende CuBr, \textbf{b} zinc blende ZnSe, \textbf{c} zinc blende GaAs, \textbf{d} rock-salt AgBr. Color indicates the contribution from anion $p$ and cation $d$ orbitals. Positive and negative -pCOHP indicate the bonding and antibonding interactions between cation and anion, respectively.
}
	\label{band}
\end{figure} 
%
\subsection{Design strategy I: antibonding from $p$-$d$ hybridization} 
To understand the unusually low $\kappa_{\rm L}$ identified in the Copper-based compounds from chemical bonding, we compare and analyze the electronic structures of three representative zinc blende compounds, namely CuBr, ZnSe, and GaAs, which show drastically different $\kappa_{\rm L}$ with values of 1.25, 19, and 45 Wm$^{-1}$K$^{-1}$ at 300 K, respectively. One particular reason of choosing these three compounds is that they have nearly the same $\overline{M}$ ($\approx$ 72 amu.), thus ensuring that the vastly different $\kappa_{\rm L}$ is not simply due to atomic mass. In Fig.\ref{band}, we show their orbital-projected band structures, the crystal orbital Hamilton population (COHP), and the corresponding molecular orbital diagrams, respectively. These three compounds have similar band structures, where the three-fold degenerate valence band maximum (VBM) is located at the $\Gamma$ point. However, the character of these bands are very different: the VBM of CuBr has more $d$-orbital contribution than ZnSe and GaAs has no $d$-orbitals at all. This can be explained by the $p$-$d$ hybridization discovered by Jaffe et al.~\cite{PhysRevB.29.1882} and Wei et al.\cite{PhysRevB.37.8958}, which states that the filled 3$d$ orbitals of Zn$^{2+}$ and Cu$^{+}$ ($d^{10}$ configuration) can participate in bonding with the 4$p$ orbitals of Se and Br in the zinc blende ZnSe and CuBr, due to the inversion-symmetry breaking of ZnSe$_4$ tetrahedra (or in general $MX_4$ tetrahedra, where $M$ and $X$ are cation and anion, respectively). In contrast, the $p$-$d$ coupling is prohibited in the case of $MX_6$ octahedra as in rocksalt structure because the odd ($d$) and even ($p$) angular momentum can't mix in the presence of inversion symmetry\cite{PhysRevB.29.1882,PhysRevB.37.8958}. The hybridization between the filled $M$-$d$ and $X$-$p$ orbitals forms $p$-$d$ bonding and $p$-$d^*$ antibonding states below the Fermi level. Importantly, the occupation of the $p$-$d^*$ antibonding states would naturally destabilize the $M$-$X$ bond strength. They further show that the overlap (strength of hybridization) between $M$-$d$ and $X$-$p$ orbitals depends on the energy difference of the constituting atoms' orbitals, provided that the $p$-$d$ coupling is allowed by the symmetry.

Based on these findings, we examine the trend of chemical bonding from CuBr to ZnSe and GaAs, wherein the energy difference between $M$-3$d$ and $X$-4$p$ orbitals increases. As we can see from the orbital mixing in Fig.\ref{band}, the hybridization between $M$-$d$ and $X$-$p$ orbitals in these compounds decreases rapidly from CuBr to GaAs. Such a trend can be understood as follows: (i) in these three compounds, the $M$-3$d$ orbitals split into $t_{2g}$ and $e_g$ levels under the tetrahedral crystal field; (ii) for the cases of CuBr and ZnSe, the $t_{2g}$ orbitals hybridize with $X$-4$p$ and form bonding and antibonding states below the Fermi level, while the $e_g$ orbitals hardly interact with $X$-$4p$ and form non-bonding states, see Fig.\ref{band}a and b; (iii) since Cu-3$d$ orbitals are higher in energy than Br-4$p$, the $p$-$d$ antibonding states in CuBr are dominated by the Cu-3$d$ orbitals and the bonding states are mainly from Br-4$p$, see Fig.\ref{band}a. In contrast, Se-4$p$ orbitals have a larger contribution to the $p$-$d$ antibonding states than Zn-3$d$ in ZnSe, see Fig.\ref{band}b, due to the lower energy level of Zn-3$d$ than Se-4$p$; and (iv) in the case of GaAs, the Ga-3$d$ is too deep to interact with As-4$p$ and therefore, there is almost no hybridization between Ga-3$d$ and As-4$p$ orbitals, and the valence bands consist of the hybridization between Ga-4$s$ and As-4$p$ orbitals only, see Fig.\ref{band}c. The filled antibonding states can be explicitly seen in the COHP analysis as shown in Fig.~\ref{band}. The projected COHP (pCOHP) between cation $M$ and anion $X$ of CuBr displays a much larger antibonding peak below the Fermi level than ZnSe, while GaAs has no antibonding peak in the valence band range at all. It is worth noting that the presence of Cu-$X$ antibonding states have been reported in other compounds with Cu$X_4$ tetrahedral coordination as well, for instance, Cu$_2$Se\cite{zhang2014electronic} and Cu$_2$S\cite{woods2020wide}, where the effects of the antibonding states on electronic structures are pronounced.

Interestingly, we find that the strong $p$-$d$ coupling between the filled Cu/Ag-$d$ and anion $p$-orbitals exists in other polyhedra with different coordination numbers as well, such as $MX_6$ octahedra, $MX_3$ planar triangle, $MX_2$ linear chain, etc. This is because the symmetry restriction of $p$-$d$ mixing\cite{PhysRevB.29.1882,PhysRevB.37.8958} only holds for certain $K$-points of the first Brillouin zone. For example, the $p$-$d$ mixing in rock-salt AgBr is only strictly prohibited at the $\Gamma$ and X points where the AgBr$_6$ octahedra preserve inversion symmetry, see Fig.~\ref{band}d. In contrast, we can see strong $p$-$d$ mixing at other $K$-points of the Brillouin-zone, especially along the $\Gamma$-K and $\Gamma$-L-U lines. Similar to the zinc blende CuBr, the $p$-$d^{*}$ antibonding states are at the top of the valence bands of AgBr. Unlike CuBr, however, the valence band maximum is shifted away from the $\Gamma$ point, increasing valley degeneracy. Meanwhile, since the $M$-$X$ bond length of rock-salt is much longer than that of zinc blende, both the bonding and antibonding peaks of rock-salt are smaller than those of zinc blende, as evidenced by pCOHP in Fig.\ref{band}d. The planar triangle and linear coordination examples are presented in Fig.\textcolor{blue}{S1}. What makes Cu and Ag significantly different from other transition metal elements such as Zn and Ga is their relatively high energy of the $d$ orbitals, which essentially determines both the $p$-$d$ coupling strength since the orbitals with similar energies are more likely to interact\cite{PhysRevB.29.1882,PhysRevB.37.8958,woods2020wide}.

\begin{figure}
	\centering
    \includegraphics[width=1.0\linewidth]{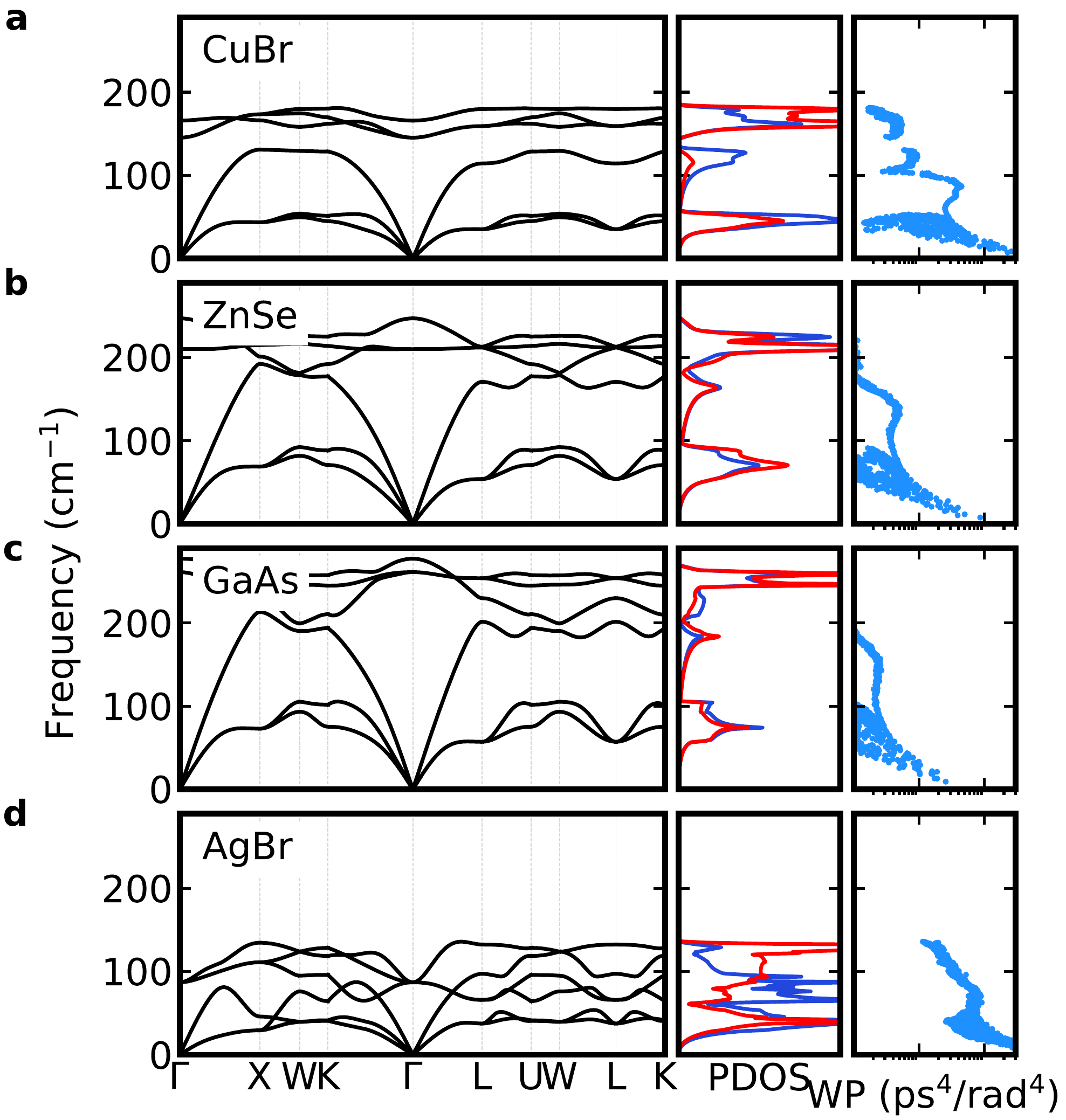}
	\caption{Phonon dispersions, phonon density of states (PDOS), and weighted phase space W of three-phonon scattering for binary compounds. \textbf{a} zinc blende CuBr, \textbf{b} zinc blende ZnSe, \textbf{c} zinc blende GaAs, and \textbf{d} rock-salt AgBr.}
	\label{phonon}
\end{figure} 

The occupation of the $d$-$p^{*}$ antibonding states is expected to weaken the $M$-$X$ bond strength, which can be quantitatively evaluated by the second-order (or harmonic) interatomic force constants. Our DFT calculations show that the 2$^{\rm nd}$IFC decreases from 17.1 eV/$\rm{\AA^2}$ in GaAs to 12.4 eV/$\rm{\AA^2}$ in ZnSe, and to 7.2 eV/$\rm{\AA^2}$ in CuBr. This trend is concomitant with the dramatic decrease of the mean sound velocity ($v_m$, GaAs: 3.24 Km/s; ZnSe: 2.50 Km/s; CuCl: 1.90 Km/s), which is an approximation of sound group velocity ($v_g$)\cite{chen2018manipulation}. As we pointed out earlier, these compounds have nearly the same $\overline{M}$, and therefore, the only reason causing a large variation in $v_m$ is the $M$-$X$ bond strength ($k$). Presumably, this is one of the main reasons that the $\kappa_{\rm L}$ of these compounds decreases significantly from GaAs to CuBr since $\kappa_{\rm L}$ is proportional to $v_m^3$\cite{toberer2011phonon,zeier2016thinking}. As shown in Fig.\ref{phonon}, in addition to the overall phonon softening (reduced phonon frequency) from GaAs to CuBr due to the reduced $M$-$X$ 2$^{\rm nd}$IFC, another remarkable feature of these phonon spectra is the flattening of the two transverse acoustic (TA) branches from GaAs to CuCl. These increasingly localized (less dispersive) acoustic phonon bands significantly increase the phonon-phonon scattering rates ($\tau^{-1}$) through providing more scattering channels, which is quantified by the weighted phase space $W$, see Fig.~\ref{phonon}. This effect has also been found in the case of an archetypal system YbFe$_4$Sb$_{12}$ exhibiting rattling and localized vibrations\cite{PhysRevB.91.144304}. It is interesting to note from the phonon density of states in Fig.\ref{phonon}a that the nearly flat TA modes are mainly from Cu in CuBr, even though the mass of Cu is smaller than Br and the lighter atom usually appears with higher vibrational frequencies. Together with the weak bonding interaction between Cu and the host framework, these flat and localized vibrations suggest that Cu exhibits rattling-like behavior in the close-packed structure. We also find the optical branches are increasingly softened around the $\Gamma$ point from GaAs to CuBr. This softening is commonly seen in the rock-salt compounds SnTe and PbTe\cite{sootsman2009new}, which has been partially attributed to the cause of low $\kappa_{\rm L}$, but it is not common in zinc blende systems\cite{delaire2011giant,PhysRevB.96.100301}. As discussed above, the bond softening mechanism is not limited to zinc blende structure. All these features observed in zinc blende structure also appear in rock-salt and AgBr ($\kappa_{\rm L}$ = 1.1 Wm$^{-1}$K$^{-1}$ at 300 K\cite{kamran2007thermal}). And since the $M$-$X$ bond length in rock-salt structure is usually longer, AgBr has much softer phonon frequencies, lower $v_m$, higher $W$, thus low $\kappa_{\rm L}$. With the above analysis and discussion in mind, we can conclude that antibonding from this unusual $p$-$d$ hybridization in Cu/Ag-based compounds can be leveraged to effectively weaken interatomic interactions, which serves as our first design strategy for the discovery of materials with intrinsically low $\kappa_{\rm L}$.

\subsection{Design strategy II: edge and face-sharing polyhedra}
In addition to forming the antibonding states, an alternative strategy to weaken the $M$-$X$ bond and hence to have lower $\kappa_{\rm L}$ is to have higher coordination number (CN)\cite{spitzer1970lattice,zeier2016thinking}. However, Cu$^{+}$ and Ag$^{+}$ are generally too small to sit in the polyhedron with CN $\ge$ 6. In this case, according to Pauling's third rule\cite{pauling1929principles}, the $M$-$X$ bond strength can be further weakened provided the $MX_4$ tetrahedra (or more generally $MX_n$ polyhedra, $n$ is the CN) are edge or even face-sharing. This is due to the following two effects: (i) the distance between two $M$ cations is significantly reduced in the edge and face-sharing polyhedron compared to that in the corner-sharing one, and (ii) the resulting Coulomb repulsion among $M$ cations is so strong that the $M$-$X$ bond has to be elongated to increase the $M$-$M$ distance. Apparently, the $M$-$X$ bond elongation is much larger in polyhedra with low CN than those with high CN because the $M$-$M$ distance is shorter in the former (Pauling's second rule\cite{pauling1929principles}). Therefore, this strategy is more effective for the tetrahedron, which is the main focus of this work. Note that although cations with a higher oxidation state have larger Coulomb repulsion, only Cu$^{+}$ and Ag$^{+}$ are considered in this work because we want to take advantage of the $M$-$X$ antibonding originating from $p$-$d$ coupling. Furthermore, both Cu$^{+}$ and Ag$^{+}$ tend to form low-coordination polyhedra, such as tetrahedra and trigonal planars with chalcogens and halogens. These anions (except Fluorine) are highly polarizable and widely used in thermoelectric materials\cite{chen2016understanding}.

\begin{figure}
	\centering
	\includegraphics[width=0.9\linewidth]{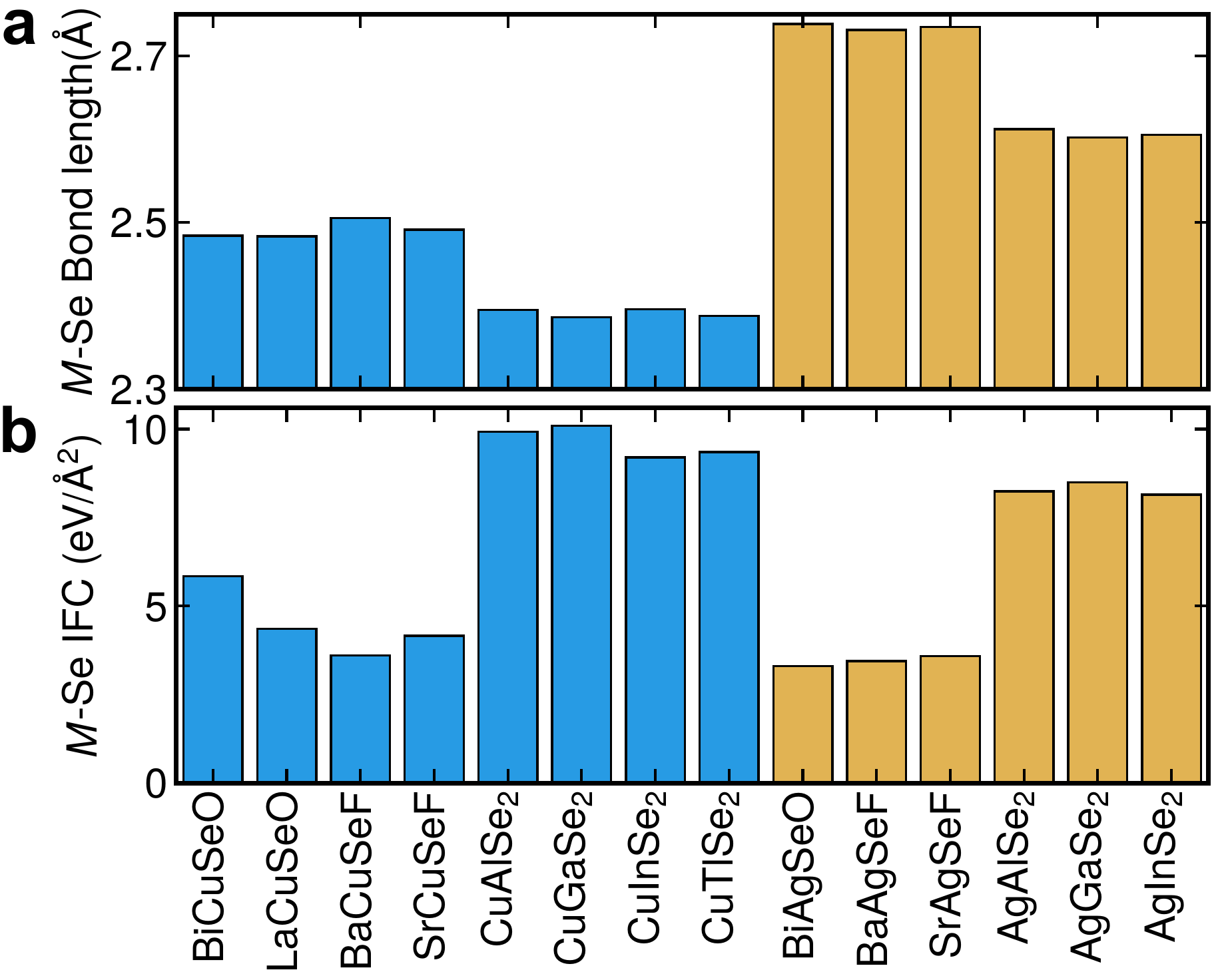}
	\caption{Bong lengths and force constants of Cu/Ag compounds with chalcopyrite and ZrCuSiAs-type structures. \textbf{a} $M$-Se ($M$=Cu and Ag) bond lengths. \textbf{b} 2$^{\rm nd}$ interatomic force constants between $M$ and Se. Blue and yellow bars are Cu$^{+}$ and Ag$^{+}$ compounds, respectively.}
	\label{IFC}
\end{figure}

\begin{figure}
	\centering
	\includegraphics[width=0.9\linewidth]{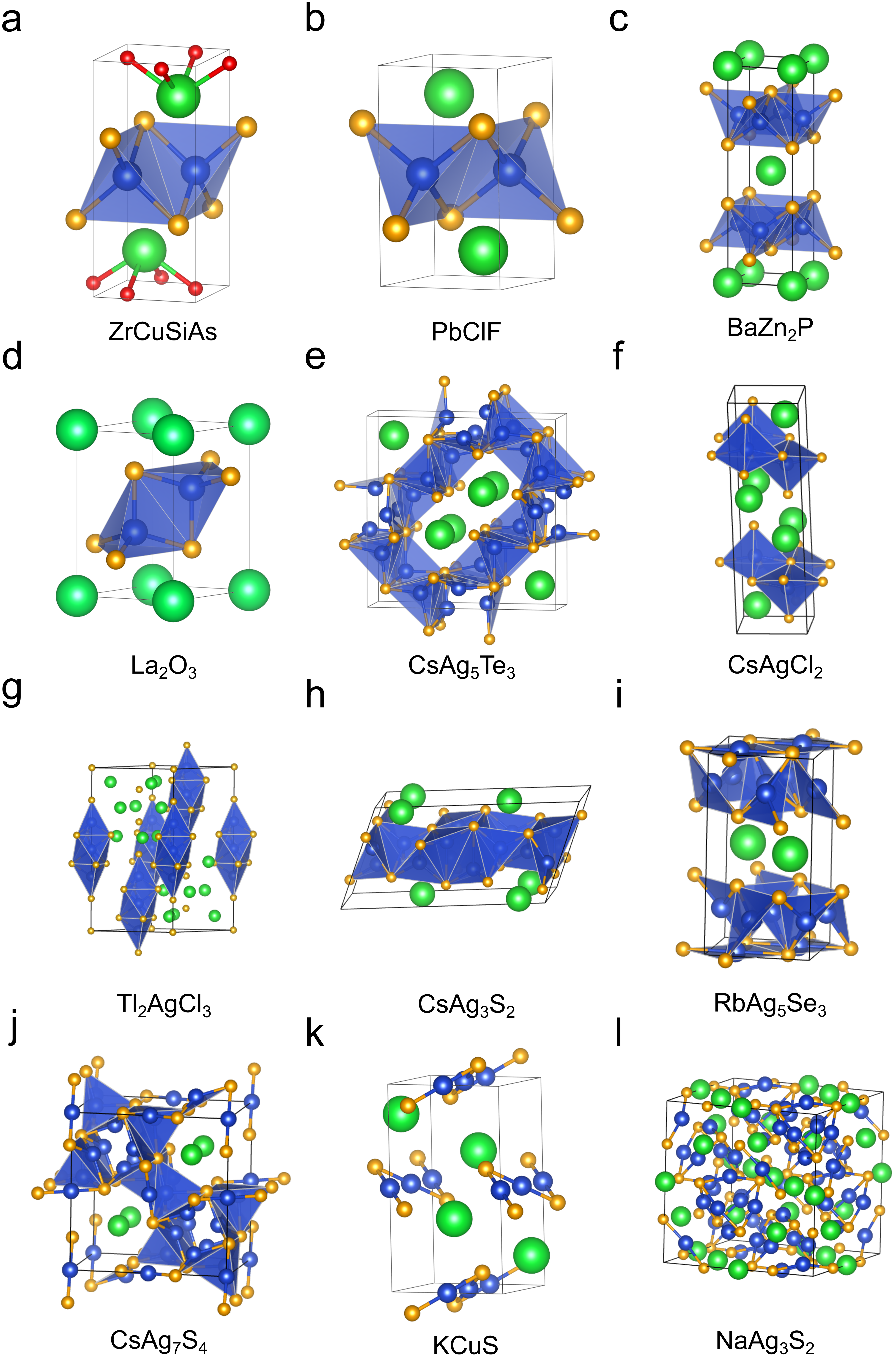}
	\caption{Prototype structures with corner, edge, or face-sharing $MX_n$ polyhedra. \textbf{a} ZrCuSiAs ($P4/nmm$), \textbf{b} PbClF ($P4/nmm$), \textbf{c} BaZn$_2$P$_2$ ($I4/mmm$), \textbf{d} La$_2$O$_3$ ($P\bar{3}m1$), \textbf{e} CsAg$_5$Te$_3$ ($P4_2/mnm$), \textbf{f} CsAgCl$_2$ ($Cmcm$), 
	\textbf{g} Tl$_2$AgCl$_3$ ($R\bar{3}$), \textbf{h} CsAg$_3$S$_2$ ($C2/m$), \textbf{i} RbAg$_5$Se$_3$ ($P4/nbm$), \textbf{j} CsAg$_7$S$_4$ ($P4/n$), \textbf{k} KCuS ($Pna2_1$), \textbf{l} NaAg$_3$S$_2$ ($Fd\bar{3}m$). Blue, green, orange, and red spheres are Cu/Ag, alkali/alkaline-earth metal, chalcogen/(Cl, Br, I), and O/F atoms, respectively.}
	\label{crystalstructure}
\end{figure}

It is worth noting that although many Cu-based compounds have been intensively optimized for thermoelectric applications\cite{Becauerel1827,qiu2016cu,zhao2010bi,zhao2014bicuseo,liu2012copper,shi2009thermoelectric,nunna2017ultrahigh,qiu2018suppression,wei2019copper}, most of these studies focus on the diamond-like structures, e.g., chalcopyrite, famatinite, and stannite, which only contain the corner-sharing Cu$X_4$ tetrahedra and are certainly not ideal thermoelectric materials in terms of $\kappa_{\rm L}$ based on our analysis. To illustrate this, in Fig.\ref{IFC} we compare the $M$-Se bond lengths and IFCs of Cu/Se and Ag/Se compounds with chalcopyrite and ZrCuSiAs-type structures. In contrast to the corner-sharing tetrahedra in chalcopyrite, the ZrCuSiAs-type structure features a layered structure with edge-sharing CuAs$_4$ tetrahedra within the layer, see Fig.\ref{crystalstructure}. It is clear that the ZrCuSiAs-type compounds (BiCuSeO, LaCuSeO, BaCuSeF, SrCuSeF, BiAgSeO, BaAgSeF, and SrAgSeF) have longer $M$-Se bond lengths and smaller IFCs than those of chalcopyrite compounds (CuAlSe$_2$, CuGaSe$_2$, CuInSe$_2$, CuTlSe$_2$, AgAlSe$_2$, AgGaSe$_2$, and AgInSe$_2$). This is also consistent with the fact that ZrCuSiAs-type compounds have lower $\kappa_{\rm L}$ than chalcopyrite, see Table~\textcolor{blue}{S1}. Another set of examples is Cu$_2X$ ($X$=S, Se, and Te), which adopts the antifluorite structure at elevated temperatures, wherein the Cu$X_4$ tetrahedra are edge-sharing in a three-dimensional (3D) network. These compounds are well-studied as thermoelectric materials due to the low $\kappa_{\rm L}$\cite{barth1926regulare,heyding1966copper,liu2012copper}. In these high-temperature phases, Cu$^{+}$ ions do not stay at the centers of Cu$X_4$ tetrahedra (the ideal position of the antifluorite structure) but hop around the center and therefore exhibit superionic behavior\cite{heyding1976crystal,oliveria1988single}. This facile ionic hopping is ultimately due to the weak Cu-$X$ bond, as the consequence of both the filled $p$-$d^{*}$ antibonding states and the 3D edge-sharing Cu$X_4$ tetrahedra. In the 3D case, since the Coulomb repulsion between $M$ cations is available in every direction, the Cu-$X$ bond is further weakened compared to the layered (two-dimensional, 2D) edge-sharing CuSe$_4$ tetrahedra, which is likely the main reason why Cu$^{+}$ migration occurs in Cu$_2X$ but not in the ZrCuSiAs-type structure.

\subsection{Screening existing compounds from ICSD}
We next apply these two design rules to search for novel, low $\kappa_{\rm L}$ compounds. We search for the experimentally synthesized compounds that have an edge or face-sharing $MX_n$ ($M$=Cu and Ag; $X$=S, Se, Te, Cl, Br, and I; $n$ $\leqslant$ 6) polyhedra from the inorganic crystal structure database (ICSD)\cite{icsd1} using the coordination environments analysis tool developed in the Materials Project\cite{Jain2013,waroquiers2017statistical}. We find 70 compounds in 21 prototype structures, as summarized in Table~\ref{kappal} and Table~\textcolor{blue}{S2}. We calculated zero-kelvin phonon spectra of all these compounds and found 27 of them have small imaginary frequencies, which means these compounds only crystallize in their reported structures at measured temperature, e.g., room temperature. Additional calculations of phonon renormalization are needed to stabilize the imaginary frequencies and compute the $\kappa_{\rm L}$. Owing to the high computational cost of phonon renormalization for such a large set of compounds\cite{xia2020particlelike}, we only calculated $\kappa_{\rm L}$ of all the compounds that have no imaginary frequency at 0 K. As shown in Fig.~\ref{crystalstructure}, the corresponding structure types can be classified in the following five categories:
\begin{itemize}
\item i. Edge-sharing polyhedra, including ZrCuSiAs, PbClF, BaZn$_2$P$_2$, La$_2$O$_3$, CsCu$_2$ICl$_2$, SrCuBiOSe$_2$, InTe-Tl$_2$Se$_2$, CsAg$_5$Te$_3$, CsAgCl$_2$, NaFeO$_2$, and Ag$_5$SbSe$_4$-type structures

\item ii. Face-sharing polyhedra, including Tl$_2$AgCl$_3$-type structure.
\item iii. Coexistence of edge-sharing and face-sharing polyhedra, including CsAg$_3$S$_2$-type structure.
\item iv. Coexistence of edge-sharing and corner-sharing polyhedra, including  RbAg$_5$Se$_3$, $\beta$-BaCu$_2$S$_2$, TiNiSi, and CsAg$_7$S$_4$-type structures.
\item v. Corner-sharing linear chain, including KCuS and Ti$_2$Ni-type structures.
\end{itemize}

\begin{table*}
	\centering	
	\caption{The properties of our calculated compounds discovered using our screening strategy from the ICSD. Prototype structure, space group, number of atoms per primitive cell ($N$), mean atom mass (amu.), the shortest $M$-$M$ distance between two $MX_n$ polyhedral ($d^{M-M}$, \AA), average sound velocity ($v_m$), and lattice thermal conductivities ($\kappa_{\rm L}$, $\kappa_{\rm L}^{xx}$, $\kappa_{\rm L}^{yy}$, $\kappa_{\rm L}^{zz}$) at 300 K of the calculated compounds. The way of polyhedral (octahedral, OC; tetrahedral, TE; trigonal, TR; trigonal bipyramidal, TB; square planar, SP) sharing with others through corner ($\cdot$), edge (-), or face ($|$) is indicated as polyhedral sharing style.} 
	\vspace{0.3cm}
	\begin{ruledtabular}
		\begin{tabular}{cccccccccc}
   Compound           &  Structure           & Polyhedral sharing        & Space group   & $N$  & $\overline{M}$& $d^{M-M}$ & $v_m$    & $\kappa_{\rm L}^{xx}$, $\kappa_{\rm L}^{yy}$, $\kappa_{\rm L}^{zz}$ \\
                      &                      & style                    &               &      &          & (\AA)     &  (m/s)   &  (WK$^{-1}$m$^{-1}$)                                                \\          
\hline
   SmCuSeO            &  ZrCuSiAs            & TE-TE                    & $P4/nmm$      &  8   &  77.22   & 2.769     &  2561   &  4.16, 4.16, 0.90                                                    \\ 
   EuCuSeO            &  ZrCuSiAs            & TE-TE                    & $P4/nmm$      &  8   &  77.62   & 2.754     &  2541   &  3.29, 3.29, 0.53                                                    \\   
   GdCuSeO            &  ZrCuSiAs            & TE-TE                    & $P4/nmm$      &  8   &  78.94   & 2.739     &  2693   &  3.69, 3.69, 0.81                                                    \\  
   DyCuSeO            &  ZrCuSiAs            & TE-TE                    & $P4/nmm$      &  8   &  80.25   & 2.713     &  2524   &  3.27, 3.27, 0.57                                                    \\    
   BiAgSeO            &  ZrCuSiAs            & TE-TE                    & $P4/nmm$      &  8   &  102.95  & 2.806     &  2245   &  1.28, 1.28, 0.22                                                    \\
   SrCuSF             &  ZrCuSiAs            & TE-TE                    & $P4/nmm$      &  8   &  50.56   & 2.761     &  3071   &  3.17, 3.17, 1.81                                                    \\   
   BaCuSF             &  ZrCuSiAs            & TE-TE                    & $P4/nmm$      &  8   &  62.99   & 2.868     &  2675   &  1.95, 1.95, 1.84                                                    \\
   SrAgSF             &  ZrCuSiAs            & TE-TE                    & $P4/nmm$      &  8   &  61.64   & 2.857     &  2698   &  1.73, 1.73, 0.45                                                    \\
   BaAgSF             &  ZrCuSiAs            & TE-TE                    & $P4/nmm$      &  8   &  74.07   & 2.977     &  2428   &  1.92, 1.92, 0.97                                                    \\
   SrCuSeF            &  ZrCuSiAs            & TE-TE                    & $P4/nmm$      &  8   &  62.28   & 2.845     &  2552   &  1.79, 1.79, 0.62                                                    \\   
   SrAgSeF            &  ZrCuSiAs            & TE-TE                    & $P4/nmm$      &  8   &  73.36   & 2.924     &  2400   &  1.67, 1.67, 0.66                                                    \\
   BaAgSeF            &  ZrCuSiAs            & TE-TE                    & $P4/nmm$      &  8   &  85.79   & 3.056     &  2199   &  1.55, 1.55, 0.85                                                    \\
   SrAgTeF            &  ZrCuSiAs            & TE-TE                    & $P4/nmm$      &  8   &  85.52   & 3.039     &  2130   &  0.97, 0.97, 0.61                                                    \\
   BaAgTeF            &  ZrCuSiAs            & TE-TE                    & $P4/nmm$      &  8   &  97.95   & 3.189     &  1775   &  0.90, 0.90, 0.30                                                    \\   
   PbCuSeF            &  ZrCuSiAs            & TE-TE                    & $P4/nmm$      &  8   &  92.18   & 2.827     &  1842   &  0.80, 0.80, 0.44                                                    \\
   PbAgSeF            &  ZrCuSiAs            & TE-TE                    & $P4/nmm$      &  8   & 103.26   & 2.892     &  1914   &  0.83, 0.83, 0.53                                                    \\
   PbAgTeF            &  ZrCuSiAs            & TE-TE                    & $P4/nmm$      &  8   & 174.11   & 2.982     &  1770   &  0.58. 0.58. 0.32                                                    \\ 
   NaCuSe             &  PbClF               & TE-TE                    & $I4/nmm$      &  6   &  55.17   & 2.848     &  2383   &  0.68, 0.68, 0.51                                                    \\          
   NaAgSe             &  PbClF               & TE-TE                    & $P4/nmm$      &  6   &  69.94   & 2.969     &  2038   &  0.74, 0.74, 0.61                                                    \\
   KAgSe              &  PbClF               & TE-TE                    & $I4/nmm$      &  6   &  75.31   & 3.141     &  1932   &  0.67, 0.67, 0.47                                                    \\ 
   CaCu$_2$S$_2$      &  La$_2$O$_3$         & TE-TE                    & $P\bar{3}m1$  &  5   &  46.26   & 2.765     &  3206   &  2.20, 2.20, 1.51                                                    \\
   BaAg$_2$S$_2$      &  La$_2$O$_3$         & TE-TE                    & $P\bar{3}m1$  &  5   &  83.44   & 3.086     &  1943   &  1.06, 1.06, 0.64                                                    \\
   TlAgI$_2$          &  InTe-Tl$_2$Se$_2$   & TE-TE                    & $I4/mcm$      &  8   & 141.51   & 3.780     &  1276   &  0.14, 0.14, 0.15                                                    \\
   CsCu$_2$ICl$_2$    &  CsCu$_2$ICl$_2$     & TE-TE                    & $P2_1/m$      &  12  &  76.30   & 2.620     &  1238   &  0.05, 0.09, 0.07                                                    \\
   AgSbTe$_2$         &  NaFeO$_2$           & OC-OC                    & $R\bar{3}m$   &  4   & 121.20   & 4.278     &  1847   &  1.34, 1.34, 0.95                                                    \\
   CsAgCl$_2$         &  CsAgCl$_2$          & TB-TB                    & $Cmcm$        &  8   &  77.92   & 4.005     &  1350   &  0.13, 0.07, 0.17                                                    \\
   TlCu$_5$Se$_3$     &  CsAg$_5$Te$_3$      & TE-TE, TE-TR             & $P4_2/mnm$    &  36  &  84.33   & 2.535     &  1842   &  0.19, 0.19, 0.34                                                    \\
   CsAg$_5$Se$_3$     &  CsAg$_5$Te$_3$      & TE-TE, TE-TR             & $P4_2/mnm$    &  36  &  101.02  & 2.838     &  1661   &  0.04, 0.04, 0.09                                                    \\             
   Tl$_2$AgCl$_3$     &  Tl$_2$AgCl$_3$      & TE$|$OC                  & $R\bar{3}$    &  18  & 103.83   & 2.996     &  1252   &  0.07, 0.07, 0.06                                                    \\
   Tl$_2$AgBr$_3$     &  Tl$_2$AgCl$_3$      & TE$|$OC                  & $R\bar{3}$    &  18  & 126.06   & 3.004     &  1176   &  0.05, 0.05, 0.04                                                    \\
   Tl$_2$AgI$_3$      &  Tl$_2$AgCl$_3$      & TE$|$OC                  & $R\bar{3}$    &  18  & 149.56   & 2.998     &  1151   &  0.04, 0.04, 0.05                                                    \\     
   KCu$_3$S$_2$       &  CsAg$_3$S$_2$       & TE-TE, TE$|$TE, TE-TR    & $C2/m$        &  12  &  48.98   & 2.563     &  2455   &  0.75, 0.61, 0.72                                                    \\ 
   RbCu$_3$S$_2$      &  CsAg$_3$S$_2$       & TE-TE, TE$|$TE, TE-TR    & $C2/m$        &  12  &  56.71   & 2.576     &  2363   &  0.46, 0.52, 0.50                                                    \\
   RbAg$_3$S$_2$      &  CsAg$_3$S$_2$       & TE-TE, TE$|$TE, TE-TR    & $C2/m$        &  12  &  78.87   & 2.845     &  1823   &  0.15, 0.21, 0.14                                                    \\
   RbAg$_3$Se$_2$     &  CsAg$_3$S$_2$       & TE-TE, TE$|$TE, TE-TR    & $C2/m$        &  12  &  94.50   & 2.849     &  1702   &  0.08, 0.12, 0.06                                                    \\
   CsAg$_3$S$_2$      &  CsAg$_3$S$_2$       & TE-TE, TE$|$TE, TE-TR    & $C2/m$        &  12  &  86.78   & 2.855     &  1759   &  0.18, 0.26, 0.17                                                    \\ 
   CsAg$_3$Se$_2$     &  CsAg$_3$S$_2$       & TE-TE, TE$|$TE, TE-TR    & $C2/m$        &  12  & 102.41   & 2.844     &  1672   &  0.08, 0.10, 0.11                                                    \\
   KAg$_3$Se$_2$      &  CsAg$_3$S$_2$       & TE-TE, TE$|$TE, TE-TR    & $C2/m$        &  12  &  86.77   & 2.849     &  1700   &  0.11, 0.12, 0.19                                                    \\        
   BaCu$_2$Te$_2$     &  BaCu$_2$S$_2$       & TE-TE, TE$\cdot$TE       & $Pnma$        &  20  &  103.92  & 2.837     &  2040   &  0.58, 1.23, 0.69                                                    \\
   AgTlSe             &  TiNiSi              & TE-TE, TE$\cdot$TE       & $Pnma$        &  12  &  130.40  & 3.887     &  1406   &  0.29, 0.46, 0.41                                                    \\
   RbAg$_5$Se$_3$     &  RbAg$_5$Se$_3$      & SP-SP, TR$\cdot$TR       & $P4/nbm$      &  18  &  95.74   & 2.997     &  1555   &  0.20, 0.20, 0.10                                                    \\
   Ag$_5$SbSe$_4$     &  Ag$_5$SbS$_4$       & TE-TE, TE$|$TE, TE$\cdot$TE, TE$\cdot$TR &$Cmc2_1$ &20&97.70& 2.831    &  1497   &  0.03, 0.01, 0.02                                                    \\
   TlCu$_7$S$_4$      &  CsAg$_7$S$_4$       & TE-TE, TE-TR, TE$\cdot$LN, TR$\cdot$LN&$P4/n$&48&  64.79   & 2.567     &  2137   &  0.05, 0.05, 0.08                                                    \\
   RbAg$_7$S$_4$      &  CsAg$_7$S$_4$       & TE-TE, TE-TR, TE$\cdot$LN, TR$\cdot$LN& $P4/n$&48& 80.74   & 2.908     &  1785   &  0.01, 0.01, 0.03                                                    \\
   CsAg$_7$S$_4$      &  CsAg$_7$S$_4$       & TE-TE, TE-TR, TE$\cdot$LN, TR$\cdot$LN& $P4/n$&48& 84.69   & 2.838     &  1756   &  0.05, 0.05, 0.08                                                    \\ 
   KCuS               &  KCuS                & LN$\cdot$LN              & $Pna2_1$      & 12   &  44.90   & 2.606     &  1579   &  1.34, 1.34, 3.12                                                    \\
	\end{tabular}
		\flushleft
	\end{ruledtabular}
	\label{kappal}
\end{table*}

\vspace{0.5cm}
\noindent \textbf{i. Edge-sharing polyhedra.} As mentioned above, $MX_4$ tetrahedra are edge-sharing with each other within the layer perpendicular to the $c$-axis in the ZrCuSiAs-type structure, see Figure~\ref{crystalstructure}a. A well-studied thermoelectric material having this structure type is BiCuSeO, which has $\kappa_{\rm L}$ of $\sim$ 1 Wm$^{-1}$K$^{-1}$ at 300 K\cite{zhao2010bi,zhao2014bicuseo}. The low $\kappa_{\rm L}$ of BiCuSeO was initially attributed to the strong anharmonicity caused by the lone-pair electrons cation Bi$^{3+}$\cite{pei2013high}. However, more recent studies suggest that the low-frequency rattling-like phonon modes coming from Cu$^+$ play more important roles in suppressing its thermal transport\cite{vaqueiro2015role,fan2017understanding}. This proposal is supported by the large atomic displacement parameter (ADP) of Cu$^{+}$ cation\cite{vaqueiro2015role}. The large ADP of Cu is due to the weak Cu-Se bond, which originates from the antibonding $p$-$d^*$ states and the edge-sharing tetrahedra. Our local coordination environment search identifies 22 other Cu and Ag-based ZrCuSiAs-type compounds, and 17 of them have no imaginary frequency. Our anharmonic phonon calculations show that all of them have relatively low $v_{m}$ and $\kappa_{\rm L}$ in terms of their $\overline{M}$, even though many of them don't have lone-pair electrons cations and all of them contain strong ionic bond $A$-O/F ($A$ is another metal element), see Table~\ref{kappal}.

The PbClF-type structure is the ZrCuSiAs-type structure with an anion or cation vacancy, and we only focus on the anion vacancy case, see Figure~\ref{crystalstructure}b. Only two Cu and one Ag compounds are reported in the ICSD, and our calculations show that NaCuTe has imaginary frequencies (see Table~\textcolor{blue}{S1}), and both NaCuSe and KAgSe have very low $\kappa_{\rm L}$ at 300 K. We note the $\kappa_{\rm L}$ of these two compounds are lower than many ZrCuSiAs-type compounds with similar $\overline{M}$. Presumably, this is because the other cation (e.g., Na and K) is also less bonded, further weakening the bonding interactions within these compounds. This can be verified by monitoring the difference of $v_m$ between these two structures. For example, although KAgSe has similar $\overline{M}$ ($\sim$ 74 aum.) with SrAgSeF, the $v_m$ of KAgSe (1993 m/s) is much smaller than that of SrAgSeF (2415 m/s).

The BaZn$_2$P$_2$-type structure has two edge-sharing $MX_4$ tetrahedra layers separated by a Ba$^{2+}$ layer, see Figure~\ref{crystalstructure}c, which is somewhat similar to the PbClF-type structure. The $\beta$ phase of BaCu$_2$S$_2$ and BaCu$_2$Se$_2$ are the only atom-ordered Cu$^{+}$/Ag$^{+}$ compounds (i.e., without partial occupancy) reported in the ICSD. Their $\kappa_{\rm L}$ have been studied experimentally\cite{li2015bacu,kurosaki2005thermoelectric}. Our calculated $\kappa_{\rm L}$ of these two compounds are $\sim$ 2.08 and 0.87 Wm$^{-1}$K$^{-1}$ (see Talbe~\textcolor{blue}{S1}), which are slightly higher than the experimental values. Note that our calculated $\kappa_{\rm L}$ are possibly overestimated for many of these compounds because our calculations do not include the higher-order phonon scattering, phonon softening due to lattice thermal expansion, and defects and grain boundary scattering. The low $\kappa_{\rm L}$ of the compounds that do not contain lone-pair-electrons cations highlight the key structural character of suppressing heat transfer from the edge-sharing Cu$X_4$ tetrahedra.

Other structures that only have edge-sharing $MX_4$ tetrahedra are La$_2$O$_3$ (CaCu$_2$S$_2$ and BaAg$_2$S$_2$, see Figure~\ref{crystalstructure}d), InTe-Tl$_2$Se$_2$ (TlAgI$_2$), CsCu$_2$ICl$_2$ (CsCu$_2$ICl$_2$), and SrCuBiOSe$_2$-type (SrCuBiOSe$_2$) structures, see Table~\ref{kappal} and Table~\textcolor{blue}{S1}. Both CaCu$_2$S$_2$ and BaAg$_2$S$_2$ have very low $\kappa_{\rm L}$ despite their small unit cell, light $\overline{M}$, and high symmetry. The other three compounds have relatively larger unit cell and low symmetry. They all have very low $\kappa_{\rm L}$. The $\kappa_{\rm L}$ of SrCuBiOSe$_2$ has been studied experimentally\cite{luo2020intrinsically}, which agrees well with our calculated values, see Table~\textcolor{blue}{S1}.


The CsAg$_5$Te$_3$-type structure contains edge-sharing AgTe$_4$ tetrahedra and edge-sharing polyhedra between AgTe$_4$ tetrahedra and AgTe$_3$ pyramids, Figure~\ref{crystalstructure}e. The distance between the Ag cations in two edge-sharing tetrahedra is 3.1 \AA\, and that between tetrahedron and pyramid is 2.9 \AA. Since this structure has a relatively high symmetry (space group $P4_2/mnm$), the complex local coordination leads to a large unit cell (36 atoms). Three compounds (TlCu$_5$Se$_3$, CsAg$_5$Se$_3$, and CsAg$_5$Te$_3$) are found in this structure type and all have low $v_m$ and ultralow $\kappa_{\rm L}$, see Table~\ref{kappal} and Table~\textcolor{blue}{S1}. Experimentally, CsAg$_5$Te$_3$ is found to exhibit ultralow $\kappa_{\rm L}$ (0.18 Wm$^{-1}$K$^{-1}$) at room temperature\cite{lin2016concerted}. Our calculated value is much lower than the experimental value, indicating that the possible breakdown of the phonon gas model\cite{PhysRevLett.125.085901}, which is evidenced by the glasslike and nearly temperature-independent $\kappa_{\rm L}$ observed in the experiment\cite{lin2016concerted}. Nevertheless, our calculation correctly predicts that CsAg$_5$Te$_3$ has an exceptionally low $\kappa_{\rm L}$.

CsAgCl$_2$ is the only structure that has edge-sharing trigonal bipyramidal AgCl$_5$, see Figure~\ref{crystalstructure}f. Since one Ag$^{+}$ coordinates with five Cl$^{-}$, the distance between Ag$^{+}$ cations in two trigonal bipyramid (4.0 \AA) is larger than that in the edge-sharing tetrahedron($\sim$ 3.0 \AA) but smaller than the edge-sharing octahedron (4.3 \AA), see Table~\ref{kappal}. The small $\overline{M}$ (78 amu.) but low $v_m$ ($\sim$ 1200 m/s) indicate the weak bonding interaction within the compound. Even though the unit cell is relatively small (8 atoms in a unit cell), such low $v_m$ still leads to an ultralow $\kappa_{\rm L}$ ($\sim$ 0.1 Wm$^{-1}$K$^{-1}$), see Table~\ref{kappal}.

NaFeO$_2$-type AgSbTe$_2$ is the only compound that has edge-sharing octahedra. In the hexagonal structure ($R\bar{3}m$ space group), AgTe$_6$ octahedra are edge-sharing and form a layer that is separated by the SbTe$_6$ octahedral layer. In the AgTe$_6$ octahedra, the Ag-Te bond (2.931 \AA) is much longer than that in the tetrahedral coordination, such as AgInTe$_2$ (2.743 \AA). The weak bond makes AgSbTe$_2$ have a relatively lower $v_m$ (1885 m/s) and $\kappa_{\rm L}$ (1.21 Wm$^{-1}$K$^{-1}$) than PbTe ($v_m$ = 2206 m/s; $\kappa_{\rm L}$ = 2.5 Wm$^{-1}$K$^{-1}$), which is also octahedrally coordinated and has larger $\overline{M}$.

Ag$_5$SbSe$_4$ is a unique structure that contains face-sharing tetrahedra, corner-sharing tetrahedra, and corner-sharing polyhedra between tetrahedra and trigonal planars, which leads to a highly distorted structure with 20 atoms per unit cell. The shortest Ag-Ag distance is from two corner-sharing tetrahedra (2.8 \AA), which is very unusual. Besides, the face-sharing tetrahedra also lead to short Ag-Ag distance (3.0 \AA). Therefore, the $v_m$ and $\kappa_{\rm L}$ are quite small in terms of $\overline{M}$ and unit cell size.

\vspace{0.5cm}
\noindent \textbf{ii. Face-sharing polyhedra.} As indicated by our design strategy II, face-sharing polyhedra are even more highly desired to lower $\kappa_{\rm L}$. The Tl$_2$AgCl$_3$-type structure has AgCl$_6$ octahedra face-sharing with two AgCl$_4$ tetrahedra at two opposite faces, Figure~\ref{crystalstructure}g. The Ag-Ag distance between the Ag in tetrahedra and the Ag in octahedra is only about 3.0 \AA, showing a strong Coulomb repulsion. However, the face-sharing polyhedra do not connect with others, limiting the effect of bond weakening by face-sharing. All the three reported compounds (Tl$_2$AgCl$_3$, Tl$_2$AgBr$_3$, and Tl$_2$AgI$_3$) contain the heavy element Tl and very ionic halogen elements. Therefore, these compounds have very low $v_m$ ($\sim$ 1200 m/s) and nearly vanishing $\kappa_{\rm L}$ from our calculations. These $\kappa_{\rm L}$ values are likely underestimated since the conventional phonon picture might break down here\cite{mukhopadhyay2018two}, which more likely happens in compounds with ultralow $\kappa_{\rm L}$.

\vspace{0.5cm}
\noindent \textbf{iii. Coexistence of edge-sharing and face-sharing polyhedra.} 
CsAg$_3$S$_2$-type structure has quite low symmetry (space group $C2/m$) due to the complex local coordination: edge-sharing AgS$_4$ tetrahedra, face-sharing AgS$_4$ tetrahedra, and edge-sharing polyhedra between AgS$_4$ tetrahedra and AgS$_3$ trigonal pyramidal, see Figure~\ref{crystalstructure}h. The shortest distance between Ag$^{+}$ cations in two face-sharing tetrahedra is only 2.84 \AA. In addition to the short Ag-Ag distance between two edge-sharing AgS$_4$ tetrahedra, the Ag$^{+}$ cation in the trigonal planar suffers from the Coulomb repulsion from the Ag$^{+}$ cations in two AgS$_4$ tetrahedra (Ag$^{+}$-Ag$^{+}$ distance is just 2.98 \AA, which is much smaller than that of 3.85 \AA\, in chalcopyrite AgGaS$_2$). Therefore, all 7 dynamically stable compounds (5 compounds have imaginary frequencies, see Table~\textcolor{blue}{S1}) of this structure have relatively low $v_m$ and $\kappa_{\rm L}$ although they have small $\overline{M}$ and fewer atoms in the unit cell.

\vspace{0.5cm}
\noindent \textbf{iv. Coexistence of edge-sharing and corner-sharing polyhedra.} 
RbAg$_5$Se$_3$-type structure has a unique layer structure consisting of corner-sharing trigonal pyramid AgSe$_3$ and edge-sharing square planar AgSe$_4$, see Figure~\ref{crystalstructure}i. Four AgSe$_3$ trigonal pyramids share with a Se$^{2-}$ corner in a layer. Four Ag$^{+}$ cations in four square planars share a Se$^{2-}$ in another layer and every two of them share a Se-Se edge. Therefore, Ag and Se form a layer. Two Ag-Se layers are separated by a Rb layer. The Ag-Ag distances between two trigonal pyramids and two square planars are 3.0 and 4.2 \AA, respectively. The short distance of two Ag$^{+}$ cations elongates the Ag-Se bond lengths. Therefore, the moderate $\overline{M}$ (95.7 amu.) can lead to relatively low $v_m$ (1531 m/s) and low $\kappa_{\rm L}$ ($\sim$ 0.2 Wm$^{-1}$K$^{-1}$), see Table~\ref{kappal}.

Different from the $\beta$-BaCu$_2$S$_2$ (BaZn$_2$P$_2$-type structure), $\alpha$-BaCu$_2$S$_2$ has both corner-sharing and edge-sharing CuSe$_4$ tetrahedra and lower symmetry (space group $Pnma$). Since the corner-sharing polyhedra do not provide bond softening as that of edge/face-sharing polyhedra (design strategy II), the $\alpha$-BaCu$_2$S$_2$ has higher $\kappa_{\rm L}$ ($\sim$ 1.5 Wm$^{-1}$K$^{-1}$) than $\beta$-BaCu$_2$S$_2$ experimentally\cite{mcguire2011transport}. BaCu$_2$Te$_2$ also crystallizes in the $\alpha$-BaCu$_2$S$_2$-type structure and our calculations show it has a relatively low $\kappa_{\rm L}$, mainly benefiting from the heavier and lower electronegativity Te.

The CsAg$_7$S$_4$-type structure has relatively high symmetry (space group $P4/n$) but a large number of atoms in a unit cell, see Figure~\ref{crystalstructure}j. It has AgS$_4$ tetrahedra, AgS$_3$ trigonal planar polyhedra, and S-Ag-S linear chains. Edge-sharing polyhedra exist between two AgS$_4$ tetrahedra and between AgS$_4$ tetrahedra and AgS$_3$ trigonal planar. Also, there are corner-sharing polyhedra between AgS$_4$ tetrahedra and AgS$_3$ trigonal planar, between AgS$_4$ tetrahedra and S-Ag-S linear chain, and between AgS$_3$ trigonal planar and S-Ag-S linear chain. The shortest Ag-Ag distance (2.83 \AA) is between two edge-sharing tetrahedra. Three compounds are reported in the ICSD, and all are dynamically stable at 0 K. Although these compounds have relatively low $\overline{M}$, the weak $M$-$X$ bond and complex structure lead to low $v_m$ and $\kappa_{\rm L}$.

AgTlSe and AgTlTe crystallize in the TiNiSi-type structure, where AgSe$_4$ tetrahedra are edge and corner-sharing with other tetrahedra. Since the AgSe$_4$ tetrahedra are slightly distorted, the Ag-Ag distance between two edge-sharing tetrahedra is larger than many cases with edge-sharing tetrahedra. However, the short Tl$^{+}$-Ag$^{+}$ distance (3.28 \AA) could provide extra Coulomb repulsion to weaken Ag-$X$ and Tl-$X$ bonds. Weak Ag-$X$ bonds together with a large $\overline{M}$ originating from Tl and Te result in slow $v_m$ and low $\kappa_{\rm L}$. Our calculated $\kappa_{\rm L}$ is slightly lower than the experimental value (0.43 Wm$^{-1}$K$^{-1}$)\cite{kurosaki2007enhancement}, see Table~\textcolor{blue}{S1}.

\vspace{0.5cm}
\noindent \textbf{v. Corner-sharing linear chain.} The only way to increase $M$-$M$ Coulomb repulsion in the $X$-$M$-$X$ linear chain is to form nonlinear corner-sharing chains. The KCuS-type structure has S-Cu-S zigzag chains along the $c$-axis, see Figure~\ref{crystalstructure}k. Therefore, the shortest Cu-Cu distance in the KCuS structure is reduced to 2.61 \AA. The Cu-S bond length in KCuS is around 2.134 \AA. Although KCuS has a very small $\overline{M}$ ($\sim$ 45 amu.) and relatively small unit cell (12 atoms), it has low $v_m$ and $\kappa_{\rm L}$, due to the weak Cu-S bond and the hollow structure.

Another interesting structure is the Ti$_2$Ni-type structure, see Figure~\ref{crystalstructure}l. NaAg$_3$S$_2$ and KAg$_3$S$_2$ adopt this structure. In these compounds, six S-Ag-S linear chains form a tetragonal cluster, where very three S-Ag-S linear chains share a S$^{2-}$ corner. Our phonon calculations show many imaginary frequencies in these two compounds, and the very localized unstable phonon modes are mainly from the Ag cation, indicating the weak Ag-S bond.

\begin{figure*}
	\centering
	\includegraphics[width=1.0\linewidth]{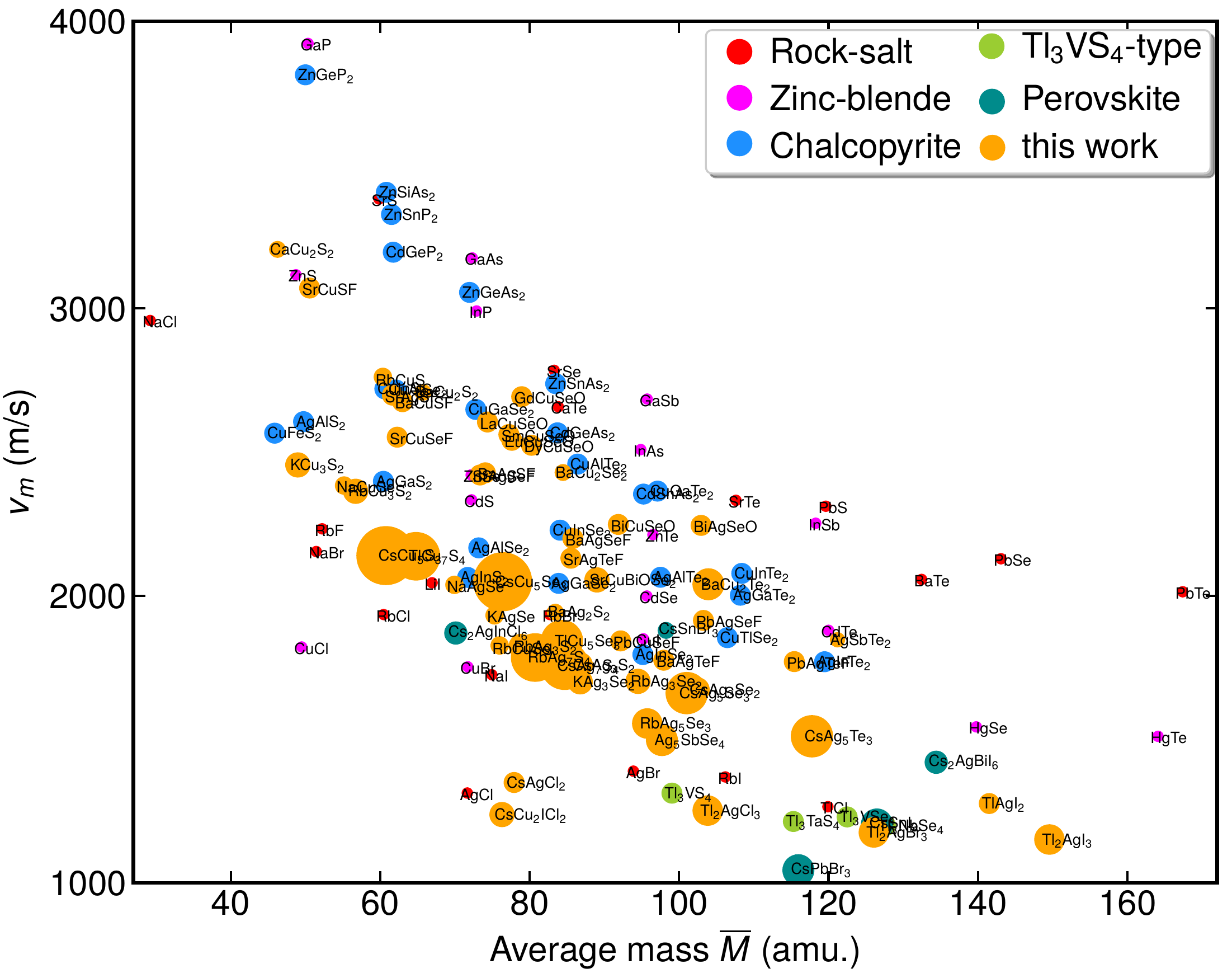}
	\caption{Dependence of $v_m$ on $\overline{M}$ for compounds with typical structures and the compounds discovered in this work. The size of the dot is proportional to the number of atoms per primitive cell.}
	\label{speedmass}
\end{figure*} 

To further illustrate the weak $M$-$X$ bonds of the compounds discovered based on our design strategies, we compare their bond strength $K$ with other compounds that have different crystal structures. Since the speed of acoustic phonon modes propagating through lattices is characterized by $v_m$, which is proportional to $\sqrt{K/M}$\cite{tritt2005thermal}, we plot the computed $v_m$ using elastic constants as a function of average mass $\overline{M}$ in Fig.~\ref{speedmass}. As expected, we observe an overall decrease of $v_m$ with increasing $\overline{M}$. As we mentioned before, the rock-salt structure has a longer cation-anion bond length and, therefore, smaller $K$ than zinc-blende. This is indeed what we see in Figure~\ref{speedmass}: rock-salt generally has lower $v_m$ than zinc-blende. Meanwhile, $K$ can also be affected by types of chemical bonds, e.g., the ionic bond is weaker than the covalent bond. We see from Figure~\ref{speedmass} that both rock-salt and zinc-blende compounds can be categorized into two groups: i. high $v_m$ group, which mainly includes the IIA-VIA and IVA-VIA rock-salt, and IIB-VIA and IIIA-VA zinc-blende; ii. low $v_m$ group, which is mainly composed of the IA-VIIA and IB-VIIA compounds. The presence of $M$-$X$ anti-bonding states is the main cause of the small $K$ in these IB-VIIA compounds, which are less ionic. The effects of the $p$-$d$ hybridization on chalcopyrite compounds can also be clearly seen in the much-reduced $v_m$ of compounds containing Cu/Ag. Overall, we see that the compounds contain Cu$^{+}$ and Ag$^{+}$ have much lower $v_m$ than the other compounds with similar $\overline{M}$. However, such an effect is less remarkable than these in the rock-salt and zinc-blende because the bond between $X$ with another cation affects the $K$ as well. It is thus conclusive that all the compounds we discovered have relatively lower $v_m$ than the other compounds with similar $\overline{M}$. Some of them even have comparable $v_m$ with the low-$v_m$ groups, halide perovskites, and Tl$_3$VS$_4$-type compounds, known as materials with ultralow $\kappa_{\rm L}$\cite{lee2017ultralow,spitzer1970lattice,mukhopadhyay2018two,xie2020all}.

\begin{figure}
	\centering
	\includegraphics[width=1.0\linewidth]{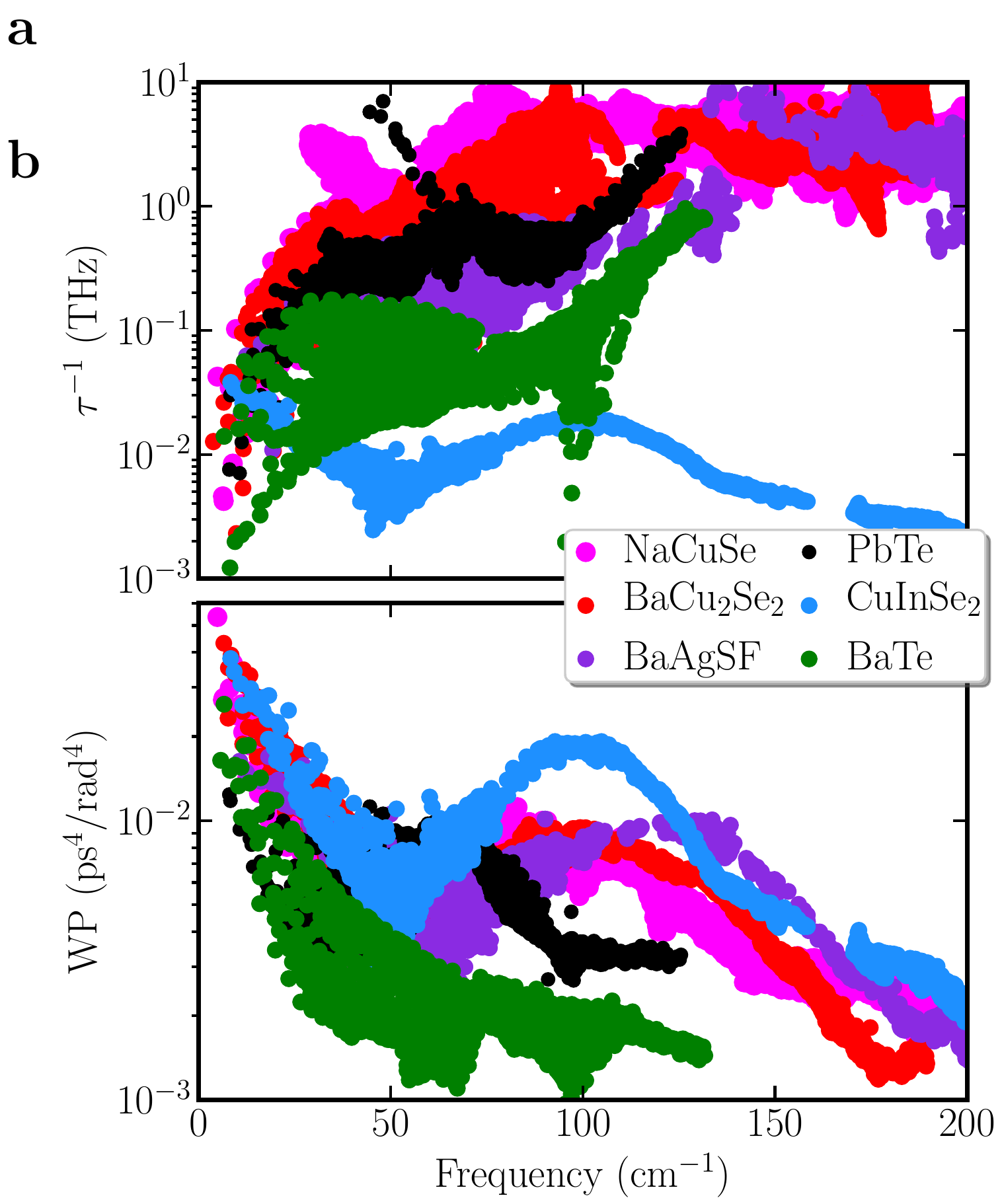}
	\caption{Anharmonic three-phonon interactions of the selected compounds. \textbf{a} phonon-phonon scattering rate ($\tau^{-1}$) as a function of phonon frequency at room temperature. \textbf{b} Weighted phase space at room temperature.}
	\label{scatter}
\end{figure} 

\begin{figure}
	\centering
	\includegraphics[width=1.0\linewidth]{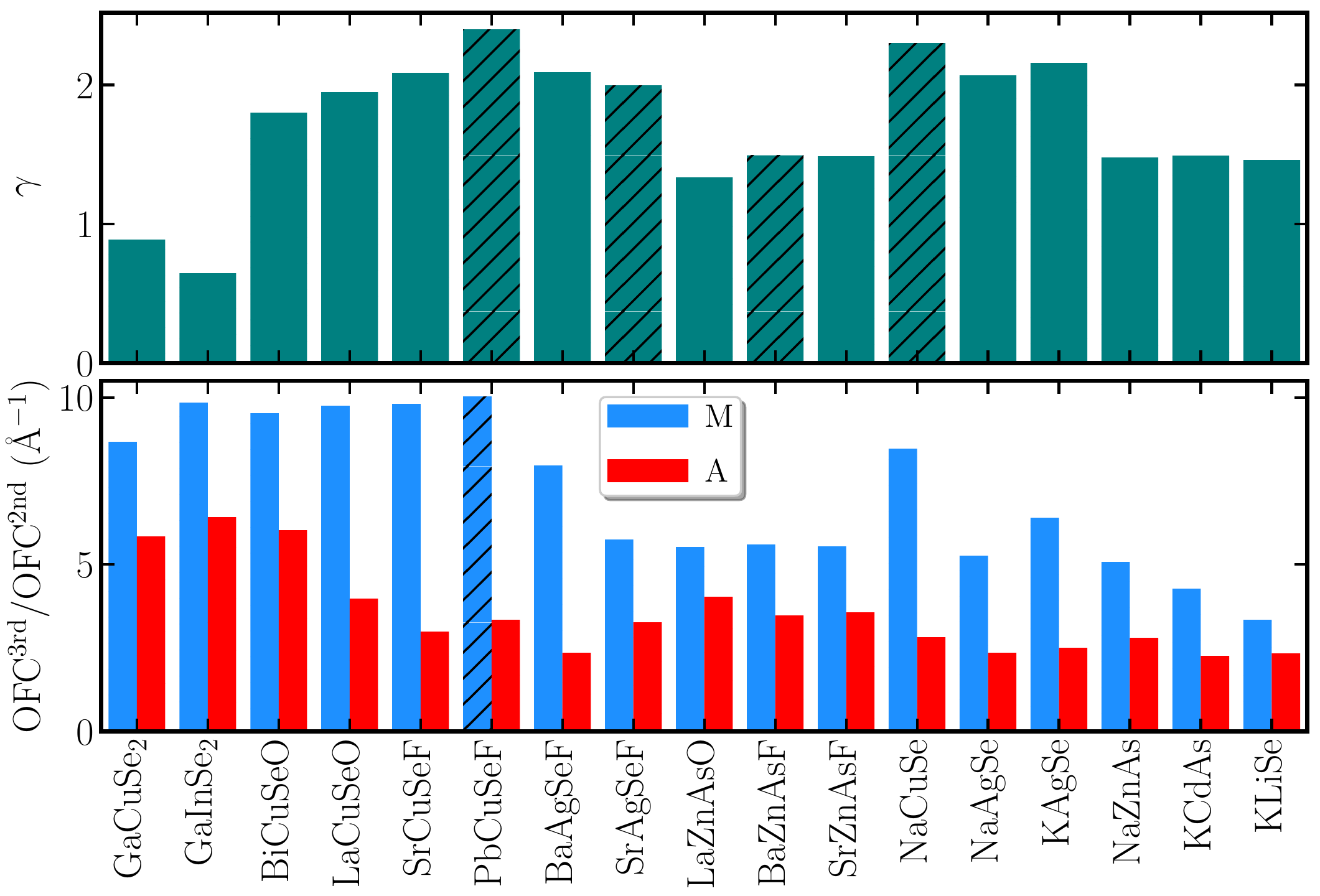}
	\caption{Gr\"uneisen parameter $\gamma$ and the ratio of third and second-order onsite force constants OFC$^{\rm 3rd}$/OFC$^{\rm 2nd}$.
	For simplicity, only the $A$ and $M$ cation sites OFC$^{\rm 3rd}$/OFC$^{\rm 2nd}$ of chalcopyrite ($AMX_2$), ZrCuSiAs ($AMXY$), and PbClF-type ($AMX$) structures are plotted. $X$ and $Y$ are anions.}
	\label{gruneisen}
\end{figure} 

In addition to lowering $v_m$, weak bond can lead to stronger anharmonicity and higher phonon-phonon scattering rate ($\tau^{-1}$). In Figure~\ref{scatter}a, we show three-phonon scattering rates $\tau^{-1}$ of six compounds with similar $v_m$ but different $\kappa_{\rm L}$: BaTe ($v_m$= 2206 m/s; $\kappa_{\rm L}$=12.6 Wm$^{-1}$K$^{-1}$), CuInSe$_2$ ($v_m$= 2318 m/s; $\kappa_{\rm L}$=6.2 Wm$^{-1}$K$^{-1}$), PbTe ($v_m$= 2206 m/s; $\kappa_{\rm L}$=2.3 Wm$^{-1}$K$^{-1}$), BaAgSF ($v_m$= 2374 m/s; $\kappa_{\rm L}$= 1.6 Wm$^{-1}$K$^{-1}$), BaCu$_2$Se$_2$ ($v_m$= 2236 m/s; $\kappa_{\rm L}$= 0.87 Wm$^{-1}$K$^{-1}$), and NaCuSe ($v_m$= 2374 m/s; $\kappa_{\rm L}$=0.6 Wm$^{-1}$K$^{-1}$). Noticeably, the Cu/Ag compounds with edge-sharing polyhedra (BaAgSF, BaCu$_2$Se$_2$, and NaCuSe) have higher $\tau^{-1}$ in the low-frequency region than chalcopyrite CuInSe$_2$, and even PbTe, a well known thermoelectric material with strong phonon anharmonicity\cite{lee2014resonant,xia2018revisiting}. In Figure~\ref{scatter}b, we show the weighted phase space (WP) of three-phonon scattering in these compounds and we can see all these compounds have rather large WP. To relate the strong scattering rates to anharmonicity, we show in Figure~\ref{gruneisen} the Gr\"uneisen parameters ($\gamma$) and the ratios of third to second-order onsite force constants (OFC$^{\rm 3rd}$/OFC$^{\rm 2nd}$) of a subset of compounds with chalcopyrite $AMX_2$ , ZrCuSiAs $AMXY$, and PbClF-type $AMX$ structures. The $\gamma$ characterizes the overall anharmonicity of a compound and OFC$^{\rm 3rd}$/OFC$^{\rm 2nd}$ indicates the bond anharmonicity associated with an ion. 
We see that high $\gamma$ values are observed in the Cu/Ag compounds with edge-sharing tetrahedra, in line with the large OFC$^{\rm 3rd}$/OFC$^{\rm 2nd}$ associated with Cu/Ag. Therefore, both low $v_m$ and high $\tau^{-1}$, which contribute to low $\kappa_{\rm L}$, are associated with the weak $M$-$X$ bond.


In order to simplify anharmonic phonon calculations, we excluded the cations with partially filled $d$ orbitals. These cations definitely can be used to destabilize the Cu/Ag-$X$ bond as well. For example, the face-sharing between AgSe$_4$ tetrahedra and CrSe$_6$ octahedra in AgCrSe$_2$ makes the Ag-Se bond so weak that the acoustic phonon modes are even dispersion-less and Ag$^{+}$ cations show superionic behavior\cite{wang2020highly,xie2020first}, and therefore it has an ultralow $\kappa_{\rm L}$\cite{wu2016revisiting}. Moreover, as we can see from Table~\ref{kappal} there are only a few compounds for many of these prototype structures. It is very likely that some of these compounds have been synthesized but not included in the ICSD, and therefore are not explored here. Meanwhile, it is still possible that some of these compounds have not been fully characterized or synthesized yet. Therefore, we expect that many low $\kappa_{\rm L}$ compounds could be discovered by decorating these structures with elements that are similar to the existing ones, which has been commonly used to predict new compounds\cite{he2020computational,pal2020accelerated}.

\begin{figure}
	\centering
	\includegraphics[width=1.0\linewidth]{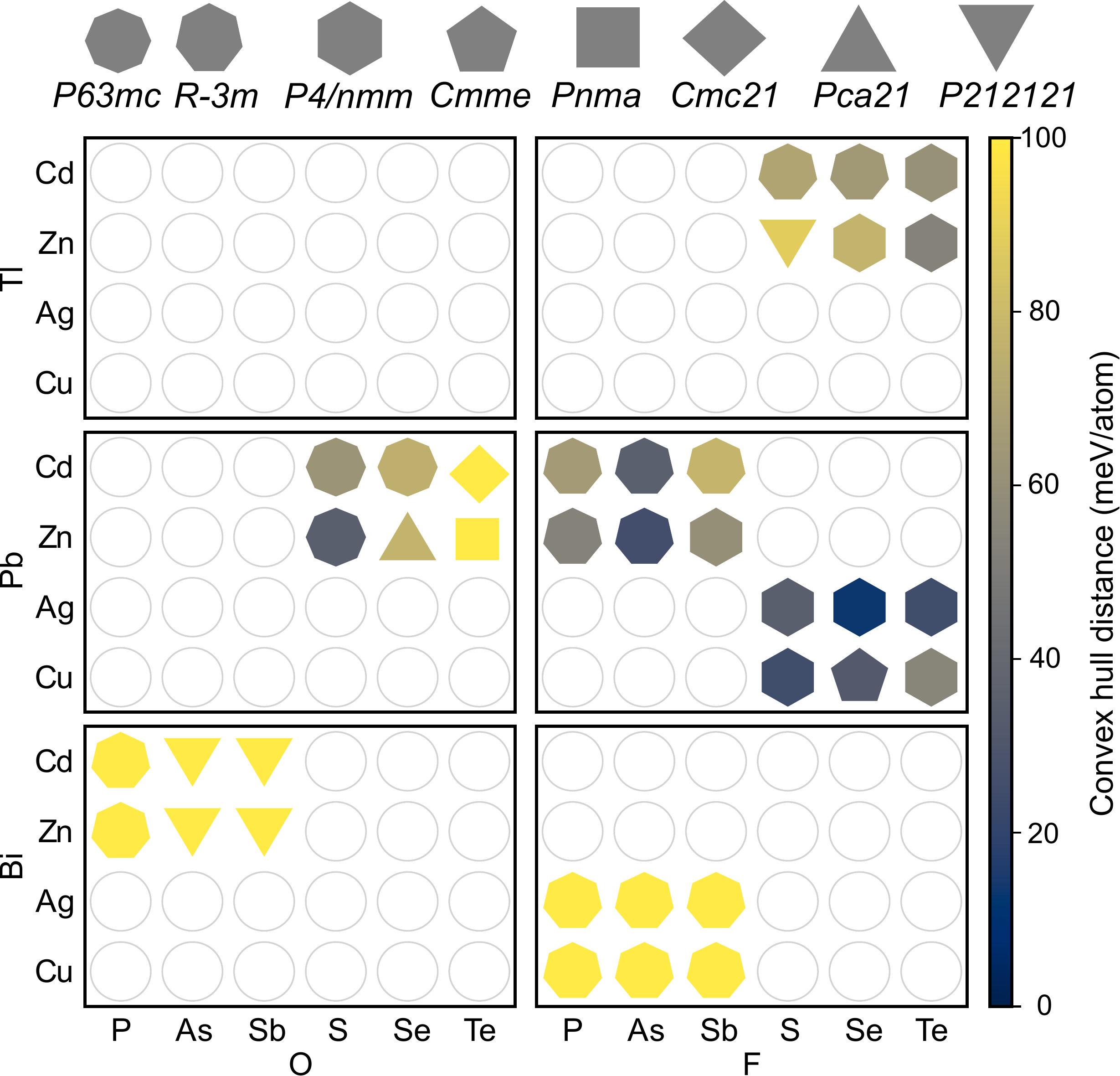}
	\caption{The heat map of the convex hull distance of $ABXY$ compounds. White unfilled circles indicate that the corresponding
		compound is not charge balanced, and therefor is not studied. Other shapes (e.g., diamond, square, triangle, \ldots) represent the different symmetries of the respective ground state structure (see the legend at the top of the figure).}
	\label{newcompounds}
\end{figure}

\begin{figure}
	\centering
	\includegraphics[width=1.0\linewidth]{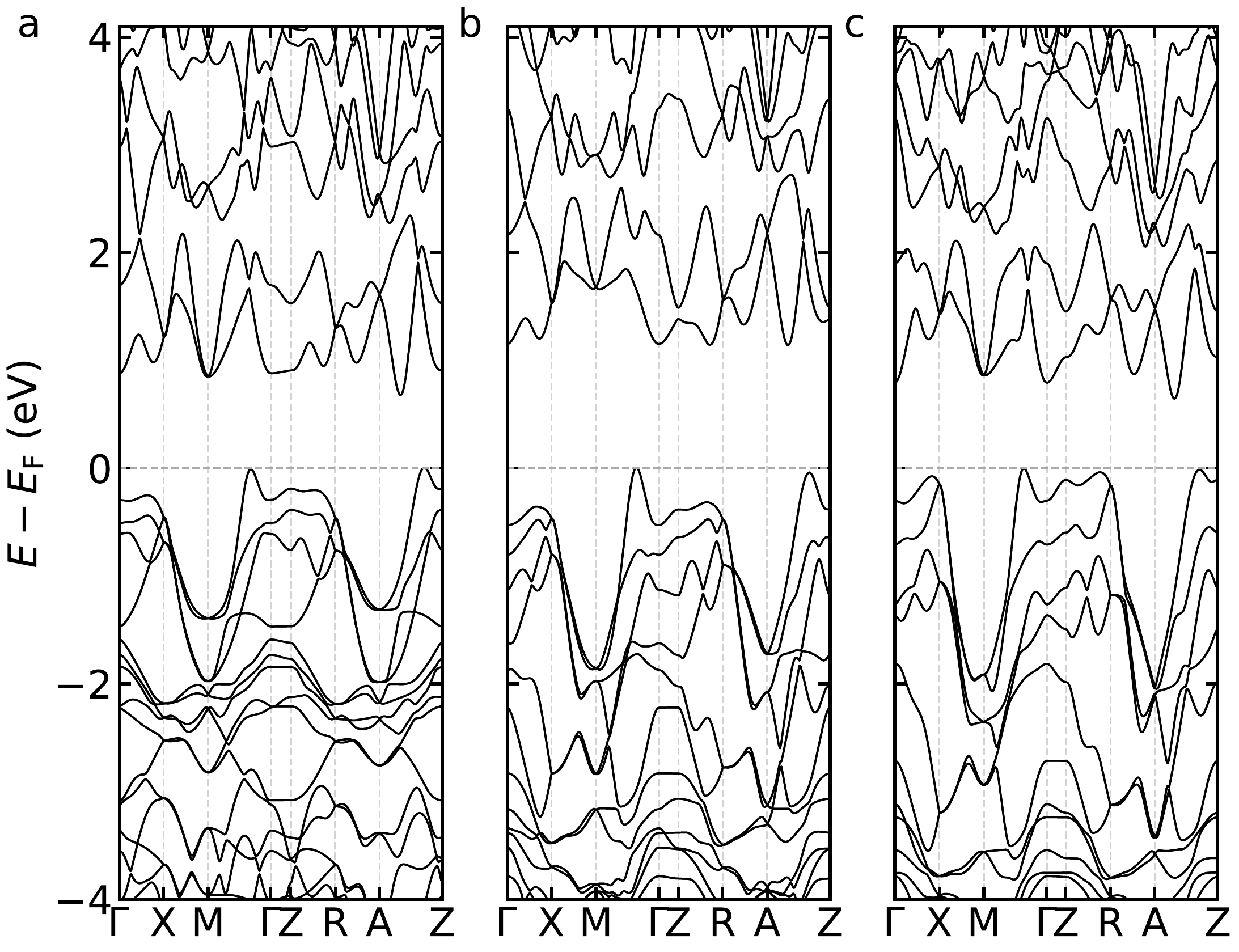}
	\caption{Electronic structures of Pb$MX$F ($M$=Cu and Ag; $X$= S, Se, and Te). (a), (b), and (c) are band structures of PbCuSeF, PbAgSeF, and PbAgTeF, respectively. The spin-orbital coupling is included.}
	\label{bandstrurenew}
\end{figure}

With our material design strategy of suppressing $\kappa_{\rm L}$, we can also rationally design as-yet synthesized thermoelectric materials by combining it with the methods focused on improving power factor. It is known that the $s$-orbital of the cations with lone-pair electrons can dramatically increase the band degeneracy of a semiconductor\cite{zeier2016thinking}, which is desired for enhancing power factors of thermoelectric materials\cite{pei2011convergence,he2019designing}.
Therefore, it is straightforward to explore new compounds by combining these two factors: edge/face-sharing $MX_{n}$ ($M$=Cu$^{+}$ and Ag$^{+}$; $X$=pnictogens, chalcogenides, and halogens) polyhedra; cations with lone-pair electrons such as Tl$^{+}$, Pb$^{2+}$, and Bi$^{3+}$. Here, we take the ZrCuSiAs-type structure as an example to illustrate such a strategy. Our DFT calculations with element-substitutions in the ZrCuSiAs-type structure and other prototype structures of the $ABXY$ composition show that the lowest energy structure of PbCuSeF, PbAgSeF, and PbAgTeF has the ZrCuSiAs-type structure, they are dynamically stable, and their formation energies are close ($\leqslant$ 25 meV/atom) to the convex hull formed by the competing phases (see Figure~\ref{newcompounds}), implying these compounds are likely synthesizable under the appropriate conditions. Further calculations reveal these compounds have relatively low $\kappa_{\rm L}$ ($<$ 1 Wm$^{-1}$K$^{-1}$). The band structures of these compounds are very similar to each other, see Figure~\ref{bandstrurenew}. Both the valence band maximum (VBM) and conduction band minimum (CBM) are located in the middle of the lines between two high symmetry points (VBM: $\Sigma$ line between $\Gamma$ and M; CBM: $S$ line between A and Z), where the valley degeneracy is 4, whereas the Ba analogs such as BaCuSeF have CBM and VBM at $\Gamma$ point with valley degeneracy of 1. Also, there are other valleys with energies close to VBM or CBM, which could further increase band degeneracy if the compounds can be properly doped. These characteristics resemble what has been observed in the excellent thermoelectric material BiCuSeO and other Bi-based compounds with the ZrCuSiAs-type structure, where the VBM and CBM are mainly from Bi 6$s$ and 6$p$ orbitals, respectively\cite{zhao2014bicuseo,he2020computational}. Similar exploration can be performed for the other structure types discovered in this work, and more promising thermoelectric materials could be potentially discovered.

\vspace{1.0cm}
In summary, we provide a feasible strategy of designing and discovering materials with low lattice thermal conductivity based on chemical bonding principles. The bond strength between Cu$^{+}$/Ag$^{+}$ and anions can be significantly weakened by filling the antibonding states originated from Cu/Ag-$d$ and anion-$p$ hybridization and enhancing Coulomb repulsion among cations in edge/face-sharing polyhedra. The consequence of the weak bond is the low speed of sound and high phonon-phonon scattering rates, which contribute to low lattice thermal conductivity ultimately. This approach is then used to screen the compounds collected in the ICSD, and 30 compounds with ultralow $\kappa_{\rm L}$ are discovered from thirteen prototype structures. We further introduce an approach to designing thermoelectric materials by combining our thermal conductivity strategy with a known method of enhancing power factor. Three as-yet synthesized thermoelectric materials with low $\kappa_{\rm L}$ and high band degeneracy are discovered by combining a structure with low lattice thermal conductivity and cations with lone-pair electrons, Pb$^{2+}$. Our material design strategy of suppressing $\kappa_{\rm L}$ is universal and is straightforward to be extended to other materials as well.

\section{Methods}
In this study, all DFT calculations are performed using the Vienna {\sl ab initio} Simulation Package (VASP)~\cite{vasp1,vasp2}. The projector augmented wave (PAW~\cite{PAW-Blochl-1994,VASP-Kresse-1999}) pseudo potential, plane wave basis set, and PBEsol~\cite{PBEsol} exchange-correlation functional were used. The qmpy~\cite{OQMD1,OQMD2} framework and the Open Quantum Material Database (OQMD)~\cite{OQMD1} was used for convex hull construction.
Second-order force constants were computed by using the finite displacement method as implemented in the \texttt{phonopy} package\cite{phonopy}.
Lattice thermal conductivities were calculated by solving the Boltzmann transport equation for phonons, as implemented in the \texttt{ShengBTE} code\cite{ShengBTE_2014}, based on the second-order force constants calculated using phonopy and third-order force constants calculated using
\texttt{thirdorder.py}\cite{PhysRevB.86.174307} and compressive sensing lattice dynamics~\cite{PhysRevLett.113.185501}.

The average sound velocity $v_m$ is calculated from bulk ($B$) and shear ($G$) moduli\cite{hill1952elastic}.

$v_m=[\frac{1}{3}(\frac{1}{v_{\rm L}^3} + \frac{2}{v_{\rm T}^3})]^{-1/3}$

$v_{\rm T}=\sqrt{\frac{G}{\rho}}$

$v_{\rm L}=\sqrt{\frac{B+\frac{4}{3}G}{\rho}}$


\noindent where $v_{\rm L}$ and $v_{T}$ are longitudinal and transversal sound velocities, respectively, and $\rho$ is mass density. The $B$ and $G$ are calculated from elastic constants. The mean values of Voigt and Reuss definition are adopted in this paper. 

\hspace{0.5cm}

\noindent \textbf{Data Availability.}\\
All data are available from the corresponding authors upon reasonable request. All codes used in this work are either publicly available or available from the authors upon reasonable request.

\bibliography{ref}
%
%
\section{Acknowledgments}
The authors acknowledge support by the U.S. Department of Energy, Office of Science and Office of Basic Energy Sciences, under Award No. DE-SC0014520 and the Center for Hierarchical Materials Design (CHiMaD) and from the U.S. Department of Commerce, National Institute of Standards and Technology under Award No. 70NANB14H012.
This research used computer resources from the National Energy Research Scientific Computing Center, a DOE Office of Science User Facility supported by the Office of Science of the U.S. Department of Energy under Contract No. DE-AC02-05CH11231, the Extreme Science and Engineering Discovery Environment, which is supported by National Science Foundation grant number ACI-1548562, and the Quest high performance computing facility at Northwestern University.

\section{Author contributions}
The research was conceived and designed by J.H. J.H. performed materials screening with help from Y.Z.
J.H. and Y.X. performed lattice thermal conductivity calculations.
J.H. performed analysis with help and suggestions from Y.X., W.L., K.P., M.K., and C.W.
C.W. supervised the whole project.
All authors discussed the results contributed to writing the manuscript.

\vspace{0.5cm}
\section{Additional information}
\textbf{Supplementary Information} accompanies this paper at .\\
\noindent \textbf{Competing interests.} The authors declare no competing financial interests. \\
\noindent \textbf{Reprints and permission} information is available online at .\\
\noindent \textbf{Publisher’s note:} Springer Nature remains neutral with regard to jurisdictional claims in
published maps and institutional affiliations.
\section{Author contributions}
All authors contributed to writing and editing the paper.
\section{Corresponding author}
Correspondence to c-wolverton@northwestern.edu \& jiangang2020@gmail.com

\end{document}